\documentclass[%
 longbibliography,
superscriptaddress,
nofootinbib,
amsmath,
amssymb,
aps,
]{revtex4-1}
\usepackage[dvipsnames]{xcolor}
\usepackage{color}
\usepackage[utf8]{inputenc}
\usepackage{amsmath}
\usepackage{amssymb}
\usepackage{float}
\usepackage{graphicx}
\usepackage{enumerate}
\usepackage[colorinlistoftodos]{todonotes}
\usepackage{appendix}
\usepackage{upgreek}
\usepackage{color, colortbl}
\usepackage{etoolbox}
\usepackage{hyperref}
\usepackage{dcolumn}
\usepackage{fullpage}
\usepackage{relsize}

\hypersetup{
    colorlinks=true,
    linkcolor=blue,
    filecolor=magenta,      
    urlcolor=blue,
    pdfpagemode=FullScreen,
    }

\newcommand{\var}[1]{{#1}}
\newcommand{\bnabla}{\mathbf{\nabla}}

\definecolor{Gray}{gray}{0.9}
\definecolor{my_emerald}{RGB}{1,99,52}
\definecolor{darkseagreen}{rgb}{0.56, 0.74, 0.56}

\usepackage[inline]{trackchanges}
\addeditor{LK}
\addeditor{RA}
\addeditor{LC}

\begin{document}

%\preprint{APS/123-QED}

\title{
The influence of thermal effects on the breakup of thin films of nanometric 
thickness}

\author{R. H. Allaire}
    \email[Correspondence email address: ]{ryan.allaire@westpoint.edu}
    \affiliation{Department of Mathematical Sciences, United States Military Academy, West Point, NY, 10996}
    \affiliation{Department of Mathematical Sciences, New Jersey Institute of Technology, Newark, NJ, 07102}
    % \altaffiliation[Currently at ]{United States Military Academy, West Point, NY, 10996}
\author{L. J. Cummings}
    \email[Correspondence email address: ]{linda.cummings@njit.edu}
    \affiliation{Department of Mathematical Sciences, New Jersey Institute of Technology, Newark, NJ, 07102}
    \author{L. Kondic}
    \email[Correspondence email address: ]{kondic@njit.edu}
    \affiliation{Department of Mathematical Sciences, New Jersey Institute of Technology, Newark, NJ, 07102}

\begin{abstract}
We apply a previously developed asymptotic model (J. Fluid. Mech. 915, A133 (2021)) to study instabilities of free surface films of nanometric thickness on thermally conductive 
substrates in two and three spatial dimensions. While the specific focus 
is on metal films exposed to laser heating, the model itself applies
to any setup involving films on the nanoscale whose material parameters
are temperature-dependent. For the particular case of metal films 
heated from above, an important aspect is that the considered heating is volumetric,
since the absorption length of the applied laser pulse is comparable 
to the film thickness. In such a setup, absorption of
thermal energy and film evolution are 
closely correlated and must be considered self-consistently.  The
asymptotic model allows for a significant simplification,
which is crucial from both modeling and computational points of view, 
since it allows for asymptotically correct averaging of the temperature
over the film thickness.  We find that the properties of the thermally conductive substrate -- in particular its thickness and rate of heat loss -- play a critical role in 
controlling the film temperature and dynamics.  The film  
evolution is simulated using efficient GPU-based simulations which, 
when combined with the developed asymptotic model, allow for fully
nonlinear time-dependent simulations in large three-dimensional 
computational domains. In addition to uncovering the role of
the substrate and its properties in determining the film evolution, one
important finding is that, at least for the considered range
of material parameters, strong in-plane thermal diffusion in the film results in negligible spatial variations of temperature, and the film evolution is predominantly influenced by 
temporal variation of film viscosity and surface tension (dictated by average film temperature), as well as thermal conductivity of the substrate.
    \end{abstract}

\maketitle

\section{Introduction}
Thin film dynamics is a well-studied problem, which has been addressed extensively from modeling, computational, 
experimental and applications points of view, as described in excellent review articles~\cite{oron_rmp97,craster_rmp09}. A particular challenge 
involves modeling external effects that couple to the fluid dynamics of the film.  Some examples include
the influence of an electric field on film dynamics \cite{Tseluiko_SIAM2007,Mema2020,chappell_o'dea_2020}, 
the competition between chemical instabilities in multi-mixture liquids and their
dewetting~\cite{Frolovskaya2008,Naraigh2010,Diez2021,Allaire_JPC2021,Thiele2013}, or the effect of permeable 
underlying substrates \cite{Davis2000,Zadrazil2006}. Thermal effects have received significant
attention as well, in particular regarding the temperature dependence of material properties, as 
discussed further below. One setup where thermal effects are clearly very important involves dynamics of 
liquid metal films deposited on thermally conductive 
substrates~\cite{trice_prb07,atena09,Saeki2011,Saeki2013,font2017,Seric_pof2018}, a setup important in the context of nanotechnology \cite{rack_nano08}, electronic coatings \cite{Zhang2010}, and
photovoltaics \cite{atwater_natmat10}, to name just a few examples.  A number of experimental works have investigated the assembly mechanism of droplets that result from liquified metal films, as described in recent reviews~\cite{Makarov2016,Hughes2017,kondic_arfm_2020}. While a number of modeling and computational studies have been carried out, theoretical modeling of thermal effects coupled with evolution of a thin film whose material parameters are temperature-dependent is
a challenging problem, which still has not been fully addressed.  Development of such a model and 
of the efficient computational methods that are required for carrying out the corresponding time-dependent simulations, is the main subject of the present paper. 

We proceed with a brief and necessarily incomplete review of relevant previous work; to put this 
discussion in the context of the present paper we first discuss briefly our earlier work~\cite{allaire_jfm2021}, in which a model for a thin molten metal film evolving on a thin thermally 
conducting substrate was proposed. Working within asymptotic long wave theory (LWT), the most significant outcome was the development of a consistent model for the coupled fluid/thermal dynamics. A key finding was that to leading order, film temperature is uniform across the (thin) film depth, with spatial and temporal evolution governed by an in-plane diffusion equation with additional terms accounting for the laser heating and heat loss to the substrate. Neglect of in-plane diffusion in the film (an approach taken in some previous works~\cite{trice_prb07,Dong_prf16,Seric_pof2018}) was shown to lead potentially to inaccurate results for heat transport, and shorter liquid lifetimes. A second focus was the influence of temperature-dependent surface tension and viscosity on the dewetting of the films. Regarding surface tension, it was found that, at least for liquid metals, the spatial variation of surface tension (Marangoni effect) did not influence the dynamics in any relevant manner. 
Temporal dependence of surface tension (via average film temperature) was found to play a much more relevant role.  Similarly, while it was found that temperature-dependent viscosity is crucial for accurately simulating films that dewet while in the liquid phase, once again temporal variation turned out to be much more relevant than spatial variation. It should be pointed out that although the dynamics of the film was coupled self-consistently to 
the thermal transport (in both substrate and film), the study was limited to asymptotically thin substrates with constant thermal properties, and the influence of substrate physical characteristics on film temperature and dynamics remains to be addressed, 
especially since in practice, substrates may be much thicker than the film itself. 

Other works have considered similar setups but with a different focus. Shklyaev {\it et al.}~\cite{shklyaev12}, for example, used LWT to derive a model similar to that of Allaire {\it et al.}~\cite{allaire_jfm2021}, but omitting laser heating, and with the underlying substrate 
(due to the assumed difference in thermal conductivities of substrate and film) modeled simply by a constant temperature gradient.  Batson {\it et al.}~\cite{batson2019} found that self-consistently 
solving for substrate temperature is crucial for the development of oscillatory free surface film instabilities, which have been previously observed (for example, when thermocapillary effects are present in multi-layer film configurations~\cite{Beerman2007} and when the film is heated from below by a substrate of sufficiently low 
thermal conductivity~\cite{shklyaev12}). Atena \& Khenner~\cite{atena09} proposed a model for liquid metals that accounts for heat transport in the substrate as well as laser heating in the film, but considers heat loss at the film surface to be relevant (see also Saeki {\it et al.}\cite{Saeki2011,Saeki2013} and Oron~\cite{oron00} in this context), leading to differences with our recent model~\cite{allaire_jfm2021}. In contrast, other works assume heat loss to the substrate to dominate over any free surface losses~\cite{trice_prb07,Dong_prf16}. Many authors have also investigated the significance of 
temperature-dependent material parameters, as discussed by Craster \& Matar~\cite{craster_rmp09}.  
Viscosity, for example, is often modeled by an Arrhenius dependence on temperature, an approach taken by Allaire {\it et al.}~\cite{allaire_jfm2021}, where it was found sufficient to use the spatially-averaged film temperature in the Arrhenius law.

In the present paper, our focus is on investigation of the role that the underlying substrate has on both the 
heating of the film and its free surface evolution. In particular, we focus on the role of substrate thickness, heat loss through the lower substrate boundary, and nonlinear effects due to temperature varying thermal conductivity. The thermal model developed in our earlier work~\cite{allaire_jfm2021} (asymptotically thin substrates, constant thermal 
properties) is extended to account for thick substrates characterized by temperature-dependent thermal conductivity. The model development is accompanied by novel GPU-based computations simulating dewetting of three-dimensional 
evolving molten films. Similarly, temperature variation of surface tension and viscosity are included, but Marangoni effects are neglected since these were demonstrated to be irrelevant in the present context~\cite{allaire_jfm2021}. 

The remainder of the paper is organized as follows. In Section~\ref{thicksub_sect:modeling}, we present the thin film equation governing the fluid dynamics and the extension of the thermal model developed previously \cite{allaire_jfm2021}. The main results are presented in Section~\ref{thicksub_sect:results}. In Section~\ref{thicksub_sect:numerical_schemes}, we outline the numerical scheme used to solve our models. In Section~\ref{thicksub_2D_flat_film}, we present results that highlight effects due to thermal transport only, in the absence of film evolution (the film surface is held flat and static even when above melting temperature); in particular the correlation between peak film temperatures and substrate thickness, as the heat loss from the substrate varies (via tuning the Biot number, Bi). In Section~\ref{thicksub_sect:2d_evolving_films} we consider evolving 2D films and investigate the influence of thermal effects on the film dynamics. In Section~\ref{thicksub_3D_evolving_films}, we present large-scale 3D numerical results for both film evolution and heat conduction. The main finding in both 2D and 3D is that the substrate heat loss, thickness, and thermal conductivity temperature dependence may all influence the final solidified film configuration, and depending on the relative strengths of these terms, films may either dewet fully or only partially by the time they resolidify.
Appendices~\ref{app_table}~-~\ref{app:numerics} provide additional information about material parameters, details of the model, and an extensive overview of the computational methods implemented. In Section~\ref{thicksub_sect:conclusion}, we present our conclusions and directions for future work.

\section{The Model}\label{thicksub_sect:modeling}
\begin{figure}[thb]
    \centering
    \includegraphics[width=0.75\textwidth]{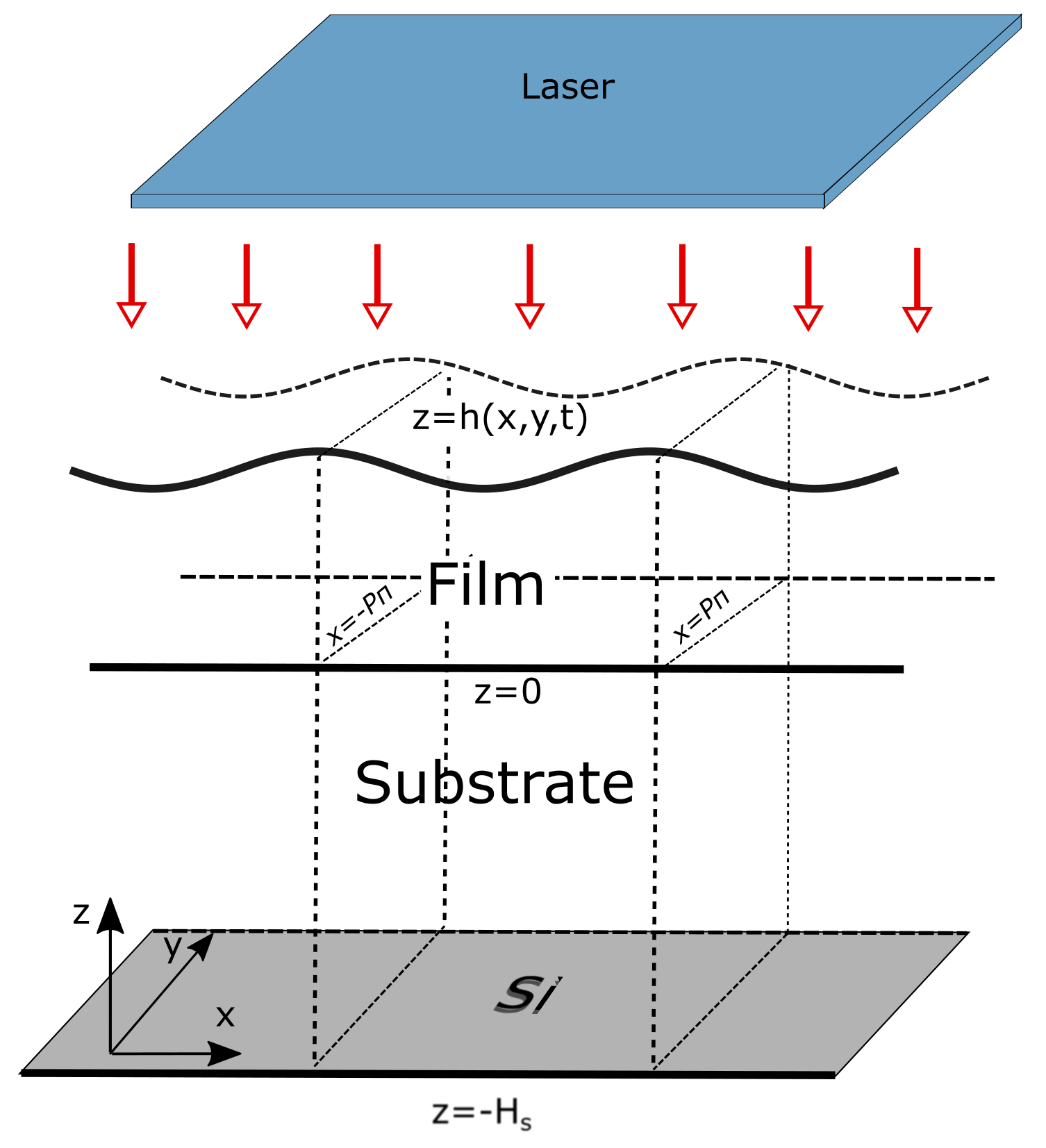}
    \caption{Schematic of a three-dimensional (3D) film with free surface $z=h(x,y,t)$, deposited on a substrate that may be much thicker than the film and is in contact with an even thicker Si slab underneath.}
    \label{thicksub_fig:schematic}
\end{figure}

Consider a free surface metal film of nanoscale thickness, $H$, and characteristic lateral length-scale $L$ (defined in terms of the wavelength of maximum growth; see Table~\ref{thicksub_table:ref_paras} and \cite{allaire_jfm2021}), which is initially solid, with air above, and in contact below (at $z=0$) with a thermally conductive solid SiO$_2$ substrate of thickness $H_s$, which may be much larger than that of the film. The whole assembly is placed upon another, thicker, slab of Si. The metal film is heated by a laser and may change phase (solid to liquid and vice-versa).  Figure~\ref{thicksub_fig:schematic} shows the basic setup. 
For later reference, Table~\ref{thicksub_table:nondim_paras_table} lists the dimensionless parameters that will be used
extensively in the paper, and the dimensional material parameters and other quantities of 
interest are specified in Appendix~\ref{app_table}, Table~\ref{thicksub_table:ref_paras}.

We define the aspect ratio of the film to be $\epsilon=H/L \ll 1$. For clarity we list a number of underlying assumptions, which 
will be discussed and where appropriate justified in the text that follows:
\begin{itemize}
    \item  the metal film evolves only when melted;  
    \item  inertial effects are negligible;
    \item phase change (melting, solidification) is fast and the associated energy gain/loss can be ignored; 
    \item  liquid-solid interactions are relevant and can be modeled by a disjoining pressure;
    \item  the laser energy is absorbed volumetrically in the film, but the substrate is optically transparent;
    \item  the film is in perfect thermal contact with the SiO$_2$ substrate at $z=0$;
    \item  heat loss in the film is only through the substrate and not through radiative losses;
%    \item  the underlying SiO$_2$ substrate is thick relative to the film, $H/H_{\rm s} \ll 1$; \note[LC]{Is this actually an assumption necessary for model validity? Here I think we should list only necessary assumptions}
    \item  the Si slab underneath the SiO$_2$ is a perfect conductor and remains at ambient temperature (this is reasonable since its thermal conductivity is much larger than that of SiO$_2$) but there is contact resistance at the interface $z=-H_{\rm s}$;
    \item  the surface tension and viscosity of the film, as well as the thermal conductivity of the substrate, may vary with temperature; and
    \item  the film does not evaporate.
\end{itemize}

With respect to the in-plane and out-of-plane length scales, $L$ and $H$ (respectively), we define in-plane coordinates $x,y$ and the out-of-plane coordinate $z$. Following Allaire {\it et al.}~\cite{allaire_jfm2021}, we choose the in-plane velocity scale  $U=\epsilon^3 \gamma_{\rm f}/(3 \mu_{\rm f})$ (where $\gamma_{\rm f}$ and $\mu_{\rm f}$ are surface tension and viscosity at melting temperature, $T_{\rm melt}$) so that the time scale, $L/U$, is comparable to the duration of the laser pulse, but the model also retains surface tension effects to leading order in $\epsilon$. Subsequently, we choose $\epsilon U$, $T_{\rm melt}$, $\mu_{\rm f} U/(\epsilon^2 L)$ and $\gamma_{\rm f}$ as the out-of-plane velocity, temperature, pressure, and surface tension scales, respectively. We take the dimensionless domain length/width to be $2 P\pi$, where $P$ is a positive integer. 

We treat the film as an incompressible Newtonian fluid, assume that the viscosity and surface tension may vary in time through the average film temperature (details to be specified below; in Appendix~\ref{thicksub_sect:spat_varying_visc} we consider spatial dependence as well), but fix material density and heat capacity at their melting temperature values. Since our focus is on substrate effects we also assume the film thermal conductivity is fixed at the melting temperature value.  However, for thick substrates, large temperature gradients could lead to significant differences in thermal conductivity across the depth. Therefore, we allow thermal conductivity of the substrate to vary with temperature and use its value at ambient temperature, $\kappa_{\rm s}$, as the thermal conductivity scale. For what follows we use $T_{\rm f}$ and $T_{\rm s}$ to denote the temperatures of the film and substrate, respectively. As will be discussed further below, to leading order (with respect to $\epsilon \ll 1$), $T_{\rm f} (x,y,t)$ is independent of the out-of-plane coordinate $z$~\cite{allaire_jfm2021}. We assume that surface tension depends linearly on average film temperature, to leading order, and is given by:
\begin{align}
    \Gamma = 1 +\frac{2{\rm Ma}}{3} (T_{\rm avg} - 1), \label{thicksub_surf_tens_dep}
\end{align}
where the Marangoni number ${\rm Ma}$ and average free surface temperature, $T_{\rm avg}(t)$, are given by
\begin{align}
    {\rm Ma} = \frac{3\gamma_{\rm T} T_{\rm melt}}{2 \gamma_{\rm f}}, \qquad     T_{\rm avg}(t)=\frac{1}{\left( 2P\pi \right)^2}\int_{-P\pi}^{P\pi} \int_{-P\pi}^{P\pi}T_{\rm f}(x,y,t)\:\mathrm{d} x \mathrm{d} y. \label{thicksub_T_bar_defn}
\end{align}
Here, $\gamma_T=(\gamma_{\rm f}/T_{\rm melt})\mathrm{d}\gamma/\mathrm{d}T_{\rm avg} \vert_{T_{\rm avg}=1}$ is the change in surface tension with temperature when the film (on average) is at melting temperature, $T_{\rm avg}=1$. For the remainder of the text we omit the argument of $T_{\rm avg}(t)$ with the understanding that it is time-dependent. More general expressions for surface tension exist that account for spatial variation of temperature (Marangoni effect); it has been shown, however, that this has little influence on film evolution in the present context and thus we omit spatial dependence of $\Gamma$ \cite{allaire_jfm2021}. 

We follow the long-wave theory approach \cite{craster_rmp09} adopted in our earlier work \cite{allaire_jfm2021}, which reduces conservation of mass and momentum to a 4th order nonlinear PDE for film thickness, $h$, written in the general form $\partial_t h+\bnabla_2\cdot\left( h \overline{\textbf{u}}\right)=0$, where $\bnabla_2=(\partial_x, \partial_y)$ is the in-plane gradient, and $\overline{\textbf{u}}=(u,v)$ is the depth-averaged in-plane fluid velocity, related to the pressure gradient. For the remaining text, vector quantities are in bold and scalar quantities are not. 
We assume that the pressure at the interface, $z=h$, obeys a modified Laplace-Young type boundary condition, which includes both free surface curvature and also liquid-solid interactions, modeled by a disjoining pressure $\Pi(h)$. 
While various forms of $\Pi(h)$ have been proposed (see~\cite{kondic_arfm_2020} for a review of this topic),  here we use
\begin{align}
     \var{\Pi}(\var{h})= \Omega \left[ \left( \frac{h_*}{h} \right)^n - \left( \frac{h_*}{h} \right)^m \right], \qquad
     \Omega = \frac{A_{\rm H} L}{6\pi \epsilon \gamma_{\rm f} h_*^3 H^3}. \label{thicksub_disjoining_pressure}
\end{align}
In Eq.~\eqref{thicksub_disjoining_pressure} the terms on the right-hand side represent the repulsive and attractive 
components, $h_*$ is the equilibrium film thickness where the attraction and repulsion balance, $A_{\rm H}$ is the 
Hamaker constant, and $n>m$ are positive exponents; in the present work, we use $(n,m)=(3,2)$ 
following Gonzalez {\it et al.}~\cite{lang13}. The thin film equation can then be written as
\begin{align}
    \partial_t h + \bnabla_2 \cdot \left[\frac{1}{\mathcal{M}} \left(  h^3  \bnabla_2 \left( \Gamma \bnabla_2^2 h + \Pi(h) \right)  \right) \right]  = 0, \label{thicksub_thin_film} 
\end{align}
where $\mathcal{M}=\mu/\mu_{\rm f}$ is the dimensionless viscosity, assumed to vary exponentially with average temperature via an Arrhenius law,
\begin{align}
    \mathcal{M} (t) &= \exp\left( \frac{E}{RT_{\rm melt}}\left( \frac{1}{T_{\rm avg}}-1 \right) \right), \label{thicksub_viscosity_eq}
\end{align}
where $R=8.314~{\rm J} {\rm K}^{-1} {\rm mol}^{-1}$ is the universal gas constant, and $E$ is the activation 
energy \cite{metals_ref_book_2004}. Other approaches have been used to implement temperature dependence of viscosity; see e.g. Kaptay~\cite{kaptay} for a comparison of Arrhenius and statistical mechanics approaches, or Oron {\it et al.}~\cite{oron_rmp97} for derivation of an analog of Eq.~\eqref{thicksub_thin_film} that includes $z$-dependence of viscosity. We follow the approach of Seric {\it et al.}~\cite{Seric_pof2018} in utilizing Eq.~\eqref{thicksub_viscosity_eq}, but we use average film temperature and thus omit spatial dependence of viscosity (shown to be irrelevant in this context~\cite{allaire_jfm2021}).

Equation \eqref{thicksub_thin_film} describes the evolution of the nanoscale thin film, which is coupled to its temperature. To determine the temperature we use an approach similar to our previous work \cite{allaire_jfm2021}, which assumed a thin substrate to allow an asymptotic reduction of the heat flow problem in both film and substrate regions. We assume (repeating some of the previously-listed assumptions for a self-contained presentation): 
(i) the film is heated volumetrically by a laser, but the SiO$_2$ substrate is transparent, (ii) heat conduction in the film is much faster than the evolution of the film, (iii) substrate heat conduction and film evolution occur on similar timescales, and (iv) film heat loss is only through the SiO$_2$ substrate, which is in perfect thermal contact with the film, and itself loses heat to an underlying Si slab of much higher thermal conductivity. To extend our previous work, we present a formulation that includes temperature-varying thermal conductivity in the substrate, $\kappa(T_{\rm s})$ %($\kappa(T_{\rm a})=1$ represents dimensionless thermal conductivity at ambient %temperature). 
(made dimensionless by scaling with $\kappa_s$, which is the substrate thermal 
conductivity at the ambient temperature, $T_{\rm a}$).  
Furthermore, we now allow the substrate to be thick, but assume negligible in-plane diffusion (in Appendix~\ref{thicksub_sect:model_validation}, we show this assumption to be valid). The leading order film temperature is found to be independent of $z$ and the model describing the transport of heat in the film/substrate system is then~\cite{allaire_jfm2021}
\begin{align}
    h \mbox{Pe}_{\rm f} \partial_t T_{\rm f} &= \nabla_2 \cdot \left( h \nabla_2 T_{\rm f} \right) - \mathcal{K} \left(\kappa(T_{\rm s})\partial_z T_{\rm s}\right)\vert_{z=0} + h \overline{Q}, \quad &&\mbox{for\ }\quad z \in \left(0,h\right), \label{thicksub_asymptotic model1} \\
    \mbox{Pe}_{\rm s}\partial_t T_{\rm s} &= \partial_z \left( \kappa(T_{\rm s}) \partial_z T_{\rm s} \right), \quad &&\mbox{for\ }\quad z \in \left(-H_{\rm s},0 \right), \label{thicksub_asymptotic_sub_eqn} \\
T_{\rm f} &= T_{\rm s}, \quad &&\mbox{on\ }\quad z=0, \label{thicksub_cont_temp_nd} \\
\kappa(T_{\rm s}) \partial_z T_{\rm s} &= {\rm Bi} \left( T_{\rm s} - T_{\rm a} \right), \quad &&\mbox{on\ }\quad z=-H_s, \label{thicksub_Bi_boundary_condition} \\
\partial_x T_{\rm f} &=0, \quad &&\mbox{on\ }\quad x=\pm P\pi, \label{thicksub_insulating_cond} \\
\partial_y T_{\rm f} &=0, \quad &&\mbox{on\ }\quad y=\pm P\pi, \label{thicksub_asymptotic_model_end}
\end{align}
where the dimensionless parameters defined by
 \begin{align}
\mbox{Pe}_{\rm f} &= \frac{\left( \rho c \right)_{\rm f} U L }{\kappa_{\rm f} }, \qquad \mbox{Pe}_{\rm s} = \frac{\left( \rho c \right)_{\rm s} U \epsilon H }{\kappa_{\rm s} }, \qquad \mathcal{K}=\frac{\kappa_{\rm s}}{\kappa_{\rm f}}\epsilon^{-2}, \qquad \mbox{Bi} = \frac{\alpha_{\rm s} H}{\kappa_{\rm s}},  \nonumber 
 \end{align}
 are the film and substrate Peclet numbers, the substrate-to-film scaled thermal conductivity ratio, and the Biot number governing heat loss from the SiO$_2$ substrate to the Si slab below, respectively.  Values for each of these parameters, as well as the film aspect ratio $\epsilon$ and the dimensionless viscosity ${\cal M}$, are given in Table~\ref{thicksub_table:nondim_paras_table}. On the right-hand side of Eq.~\eqref{thicksub_asymptotic model1} the terms, from left to right, represent lateral diffusion, film heat loss due to contact with the substrate, and the laser heat source, respectively. Equation~\eqref{thicksub_asymptotic_sub_eqn} reflects the assumption that heat flow in the substrate is affected by out-of-plane diffusion only. Since the substrate thickness may actually be comparable in size to the domain length, dropping lateral substrate diffusion is not necessarily a consequence of the leading order approximation of heat conduction in $\epsilon$, 
 but rather an assumption, justified later in  Appendix~\ref{thicksub_sect:model_validation} by showing that in-plane derivatives of substrate temperature are orders of magnitude smaller than those in the out-of-plane direction. Equation~\eqref{thicksub_cont_temp_nd} represents continuity of film/substrate temperatures and the nonlinear boundary condition in Eq.~\eqref{thicksub_Bi_boundary_condition} represents heat loss from the SiO$_2$ substrate to the underlying Si slab, assumed to be at ambient temperature, $T_{\rm a}$. Values of the heat transfer coefficient, $\alpha_{\rm s}$, in the definition of ${\rm Bi}$ are difficult to find in the literature so in this work we consider ${\rm Bi}$ to be a variable parameter within the range given in Table~\ref{thicksub_table:nondim_paras_table}. The lateral boundaries are thermally insulated, Eq.~\eqref{thicksub_insulating_cond} and \eqref{thicksub_asymptotic_model_end}. The above model assumes that radiative losses are negligible relative to heat loss to the substrate. By a simple energy argument, we find that the time scale on which radiative losses would be relevant is on the order of milliseconds, orders of magnitude longer than the time scales of the laser pulse and consequent flow considered here; see Appendix~\ref{thicksub_sect:radiative_losses} for more details. In the present 
 work we do not consider the details of the phase change process, and in particular we ignore the
 contribution of latent heat to the energy balance (such effects were considered in a similar context recently by Trice {\it et al.}~\cite{trice_prb07}, who found them to be negligible); also we assume that phase change is instantaneous, as in Seric {\it et al.}~\cite{Seric_pof2018}.  
 
 We assume the film-averaged heat source, $\overline{Q}$ in Eq.~\eqref{thicksub_asymptotic model1}, representing external volumetric heating due to the laser at normal incidence, is given by \cite{trice_prb07, Seric_pof2018},
 \begin{align}
     \overline{Q} &= \frac{1}{h}\int_0^h F(t) \left[ 1-R(h) \right] \exp{\left[-\alpha_{\rm f}\left(h-z \right)\right]} dz, \label{thicksub_Q_eqn} \\
    F(t) &= C \exp \left[ -\left(t-t_{\rm p} \right)^2/(2\sigma^2) \right], \quad C = \frac{E_0 \alpha_{\rm f} L^2}{ \sqrt{2 \pi}\sigma t_{\rm s} H \kappa_{\rm f}   T_{\rm melt}}, \nonumber
 \end{align}
where $C$ is a dimensionless constant proportional to the laser fluence, $E_0$, $\alpha_{\rm f}^{-1}$ is the (scaled) absorption length for laser radiation in the film, and $F(t)$ describes the temporal shape of the laser, taken to be Gaussian centered at $t_{\rm p}$ and of width $\sigma=t_{\rm p}/ ( 2 \sqrt{2 \ln{2}} )$. For the reflectivity of the film, $R(h)$, we use~\cite{trice_prb07,Seric_pof2018}
\begin{align}
 R(h)=r_0 \left(1-\exp \left(-\alpha_{\rm r} \var{h} \right) \right), \nonumber
\end{align}
where $r_0$ and $\alpha_{\rm r}$ are dimensionless fitting parameters, specified in Table~\ref{thicksub_table:ref_paras} in Appendix~\ref{app_table}. 

\begin{table}[H]
\centering
\setlength{\tabcolsep}{0.5em} % for the horizontal padding
{\renewcommand{\arraystretch}{1.17} % for the vertical padding
  \begin{tabular}{ | l | l | l | l |} \hline 
 \textbf{Dimensionless Numbers} & \textbf{Notation} & \textbf{Value} & \textbf{Expression}  \\ \hline 
     Aspect Ratio & $\epsilon$ & $0.347$ & $H/L$ \\
    Film Peclet Number & $\mbox{Pe}_{\rm f}$ & $1.42 \times 10^{-3}$ & $(\rho c)_{\rm f} U L/\kappa_{\rm f}$ \\
    Substrate Peclet Number & $\mbox{Pe}_{\rm s}$ & $2.17 \times 10^{-2}$ & $(\rho c)_{\rm s} U \epsilon H/\kappa_{\rm s}$ \\
    Biot Number & ${\rm Bi}$ & $10^{-3}-10^{3}$ & $\alpha_{\rm s} H/\kappa_{\rm s}$ \\
   Thermal Conductivity Ratio & $\mathcal{K}$ & $0.034$ & $\kappa_{\rm s}/(\epsilon^2 \kappa_{\rm f})$ \\ 
   Range of Dimensionless Viscosity & $\mathcal{M}$ & $0.028-1$ & $\mu/\mu_{\rm f}$ \\ \hline
 \end{tabular} }
  \caption{Dimensionless Parameters Based on Material parameters in Table~\ref{thicksub_table:ref_paras}.
  }
 \label{thicksub_table:nondim_paras_table}
\end{table}

%=======================================================================%
%=======================================================================%

\section{Results} \label{thicksub_sect:results}
After outlining our numerical approach in Section~\ref{thicksub_sect:numerical_schemes}, we  consider 2D films with free surface $z=h(x,t)$ in Section~\ref{thicksub_2D_flat_film} and Section~\ref{thicksub_sect:2d_evolving_films}, focusing on the influence of substrate thickness, Biot number, and variable substrate thermal conductivity. In Section~\ref{thicksub_3D_evolving_films} we expand our consideration to 3D films with free surface $z=h(x,y,t)$.

\subsection{Numerical schemes}\label{thicksub_sect:numerical_schemes}
In the 2D case, Eq.~\eqref{thicksub_thin_film} for $h(x,t)$ is solved using the approach of our earlier work \cite{allaire_jfm2021}, with spatial discretization commensurate with the equilibrium film thickness, $\Delta x = h_*=0.1$. Eq.~\eqref{thicksub_thin_film} can be rewritten as $\partial_t h + \partial_x J = 0$ for some flux $J$, and a Crank-Nicolson scheme is used for the time-stepping, turning Eq.~\eqref{thicksub_thin_film} into a nonlinear system of algebraic equations
\begin{align}
    \frac{h_i(t+\Delta t)- h_i(t)}{\Delta t} = \frac{1}{2} D_i(t+\Delta t) + \frac{1}{2}D_i(t), \quad i=1,2,\ldots, N, \label{thicksub_discrete_thin_film}
\end{align}
where $h_i(t) \approx h(x_i,t)$, $\{x_i\}$ is a $N$-point spatial discretization, and $D_i$ is a discretization of $\partial_x J$, at $x_i$. Although any iterative method for solving nonlinear equations would suffice to solve Eq.~\eqref{thicksub_discrete_thin_film}, we use Newton's method; since Eq.~\eqref{thicksub_discrete_thin_film} must be solved at each time-step, the rapid quadratic convergence ensures faster computing times. The initial condition takes the form of a small perturbation to a flat film $h=h_0$,
\begin{align}
    h(x,0) = h_0 \left( 1+ \delta \cos \left( x \right) \right), \label{thicksub_h_ic1}
\end{align}
where $h_0 \delta$ is the perturbation amplitude ($|\delta| \ll 1$), and the wavelength of the perturbation is equal to the domain length, $2\pi$ (see Table~\ref{thicksub_table:ref_paras} in Appendix~\ref{app_table} for the physical sizes).

A similar approach is used to solve Eq.~\eqref{thicksub_asymptotic_sub_eqn} for the substrate temperature $T_{\rm s}$, while for the film temperature $T_{\rm f}$ in Eq.~\eqref{thicksub_asymptotic model1} an implicit-explicit methodology is used (see the Appendix 
of Allaire {\it et al.}~\cite{allaire_jfm2021} for more details). The film and substrate are initially fixed at room temperature,
\begin{align}
    T_{\rm f}(x,0) = T_{\rm s}(x,z,0) = 0. 
\end{align}
During the initial laser heating both film and substrate temperatures are found by solving Eqs.~\eqref{thicksub_asymptotic model1}--\eqref{thicksub_asymptotic_sub_eqn} with the film flat and static until it melts, which we deem to happen when the minimum film temperature (over space) surpasses $T_{\rm melt}$. Film evolution, film temperature, and substrate temperature are then sequentially found at each time step. Once the minimum film temperature decreases past $T_{\rm melt}$ the film is considered solid. After this time, only film and substrate temperatures are solved for; we no longer evolve the free surface, which is frozen in what we refer to as its final configuration. 

A successful time iteration requires that two criteria are met for both film evolution and heat conduction: (i) the iterative method should converge to a relative error tolerance of $10^{-9}$ in fewer than $10$ iterations; and (ii) the relative truncation error should be less than $10^{-3}$. If either (i) or (ii) are not satisfied, the time step is decreased and the equations are integrated again.  For more details regarding the 2D numerical scheme see Appendix~\ref{thicksub_sect:2d_numerics}. 

For the 3D simulations, one needs to be careful with the choice of the initial condition, so as to produce a surface $h(x,y,0)$ with 
perturbations that are uncorrelated (in the $x$ and $y$ directions) and that excite a significant number of Fourier 
modes (note that using simply a sum or a product of sines and cosines with random amplitudes produces noise that is not random).  
Here we follow in spirit the approach of Lam {\it et al.}~\cite{Lam2019}, where the initial condition is given by
\begin{align}
    h(x,y,0) = h_0 \left( 1 + \delta \eta(x,y) \right), \label{thicksub_random_ic_eq}
\end{align}
$\eta(x,y)$ is a random perturbation, and as in the 2D case, $\delta=0.01$.

Equation \eqref{thicksub_thin_film} is written as $\partial_t h + \bnabla_2 \cdot \mathbf{J} = 0$, with flux $\mathbf{J}$, 
and solved for $h(x,y,t)$, via an alternating-direction 
implicit (ADI) method combined with the Newton iterative method described above ($D_i, h_i$ in
Eq.~\eqref{thicksub_discrete_thin_film} are now replaced by $D_{i,j},h_{i,j}$)~\cite{Lam2019}. Equation.~\eqref{thicksub_asymptotic model1} is 
now solved using an implicit-explicit ADI approach, which consists of a predictor and corrector step.
Equation~\eqref{thicksub_asymptotic_sub_eqn} is solved similarly to the 2D case, except now $T_{\rm s}=T_{\rm s}(x,y,z,t)$. 
Due to the dependence on three spatial variables, this equation alone amounts to a significant number of systems of discrete 
nonlinear equations to be solved at each time-step. Similarly, Eq.~\eqref{thicksub_thin_film} and 
Eq.~\eqref{thicksub_asymptotic model1} lead to large discrete systems, which present a daunting computational challenge. To enhance computational performance the equations are solved in parallel using the Compute Unified Device Architecture (CUDA) programming framework \cite{cuda} developed by NVIDIA\textsuperscript{\textregistered}, which utilizes graphics processing units (GPUs). 
In a similar context, Lam {\it et al.}~\cite{Lam2019} showed that GPUs offer significant computational advantages over traditional (CPU) computing, especially when large domains are considered. The parallel numerical schemes used for heat conduction are described in Appendix~\ref{thicksub_sect:3d_numerics}.

\subsection{Flat film results - influence of substrate thickness, Biot number, and thermal conductivity}\label{thicksub_2D_flat_film}
In this section we suppress dewetting in the molten film and consider the static flat film $h=h_0$, focusing on the influence of substrate properties on film temperature. In particular, we analyze the influence of (i) the substrate thickness, (ii) the substrate heat loss, and (iii) nonlinear effects due to temperature-dependent thermal conductivity in the substrate (compared with constant thermal conductivity, $\kappa=1$). For more details on the model used for the thermal conductivity, see Appendix~\ref{thicksub_sect:thermal_conduct}. In the following discussion we focus on two quantities: peak film temperature, $T_{\rm peak}$ (the maximum spatially-averaged film temperature attained by the film over the duration of the simulation), and the liquid lifetime (LL) of the film, defined as the time interval during which the average film temperature remains above melting ($T_{\rm avg}>1$).

\begin{figure}[thb!]
    \centering
    \includegraphics[width=0.8\textwidth]{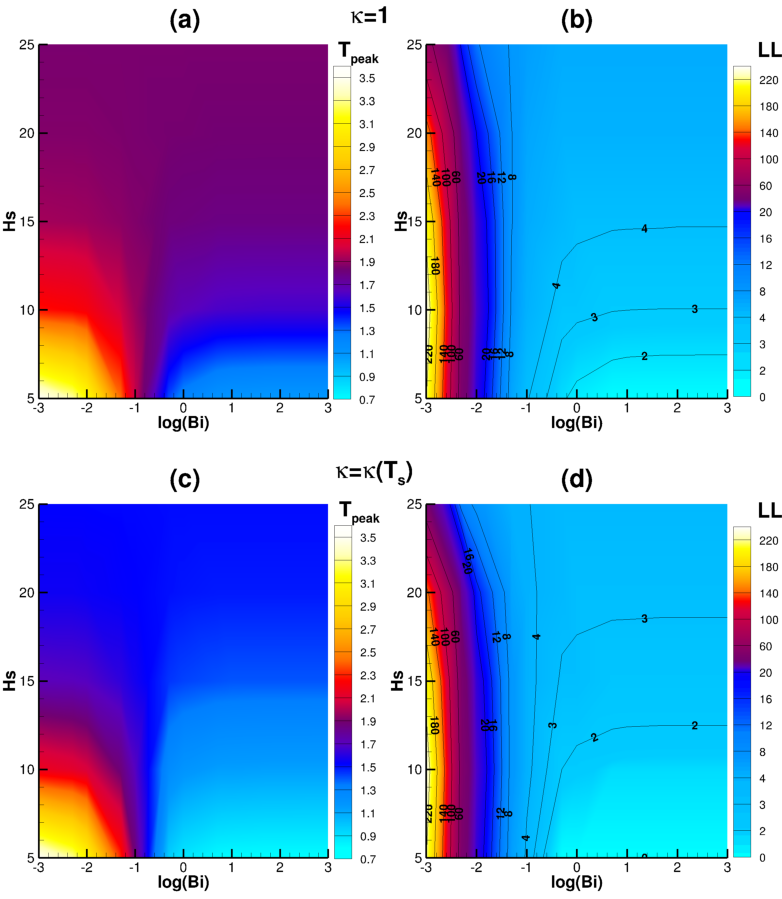}
    \caption{Phase plane plots of the film peak temperature, $T_{\rm peak}$ and liquid lifetime (LL). Here surface tension and viscosity are fixed at the melting temperature values, $\Gamma=\mathcal{M}=1$. (a, c) $T_{\rm peak}$ for thermal conductivity fixed at room temperature ($\kappa=1$), or temperature-dependent, $\kappa=\kappa(T_{\rm s})$. (b, d): corresponding results for LL. Log base 10 is used on the horizontal axes and the color bars for (b, d) are nonuniform.}
    \label{thicksub_fig:flat_film_contours}
\end{figure}

Figures \ref{thicksub_fig:flat_film_contours}(a) and (b) show phase plane plots of $T_{\rm peak}$ and LL, respectively, for various values of substrate thickness $H_{\rm s}$ and Biot numbers, ${\rm Bi}$; see Eq.~\eqref{thicksub_Bi_boundary_condition}. A zero Biot number corresponds to a perfectly insulated substrate that loses no heat to the underlying Si slab, while ${\rm Bi} \to \infty$ corresponds to a poorly insulated substrate in contact with a Si slab at ambient temperature, $T_{\rm a}$ (in Eq.~\eqref{thicksub_Bi_boundary_condition} this corresponds to a Dirichlet boundary condition, $T_{\rm f}=T_{\rm a}$). In Fig.~\ref{thicksub_fig:flat_film_contours}(a) we see that films on well-insulated substrates (Bi~$\ll 1$) retain more heat and reach higher peak temperatures than those on their poorly-insulated counterparts (Bi~$\gg 1$). In Fig.~\ref{thicksub_fig:flat_film_contours}(b) this corresponds to longer LLs for Bi~$\ll 1$. Note that here the LL scale is nonuniform and the LL varies with substrate thickness, even for ${\rm Bi}<10^{-1}$. Furthermore, we see little variation in $T_{\rm peak}$ for ${\rm Bi} \in[1,10^{3}]$, which manifests in Fig.~\ref{thicksub_fig:flat_film_contours}(b) as near-horizontal constant LL contour lines in this range, compared to those in the remaining range of ${\rm Bi}$ where LL varies significantly. Between ${\rm Bi}=10^{-1}$ and ${\rm Bi}=1$ there is a sharp transition in peak temperature and LL. This is primarily due to the changing balance between the heating of the film-plus-substrate and the heat loss from the substrate (there is perfect thermal contact at the film--substrate interface, and 
since radiative losses are neglected no heat is lost at the film's free surface). For substrates perfectly insulated from below, heat is retained in the substrate (and thus the film, due to the perfect thermal contact) more so than in the poorly-insulated case, where the film rapidly loses heat to a near-room-temperature substrate.

The influence of substrate thickness is also significant, and depends strongly on the value of ${\rm Bi}$. For well-insulated substrates (Bi~$\ll 1$), peak average film temperature decreases with increasing $H_{\rm s}$, while for poorly-insulated substrates (Bi~$\gg 1$) peak temperature increases with $H_{\rm s}$. This is again due to the competition between the absorption of heat in the substrate and the heat loss to the underlying slab at its lower boundary, $z=-H_{\rm s}$. For ${\rm Bi}\ll 1$, the thicker the substrate the more thermal energy it absorbs (due to the greater volume) and retains (due to the insulated lower boundary), leaving less heat in the film (see Supplementary materials, movie1). For ${\rm Bi} \gg 1$, substrate heat loss is rapid and the farther the interface at $z=-H_{\rm s}$ is from the molten film, the less heat is lost from the film (see Supplementary materials, movie2). Therefore, in this case thicker substrates yield higher film peak temperatures.  Liquid lifetime is, in general, positively correlated with peak temperature, despite differences in cooling. Furthermore, peak temperatures are similar for substrates thicker than $H_{\rm s}=20$ (beyond this value the substrate effectively behaves as one of infinite depth). The exact solution for a flat film on an infinite substrate $H_{\rm s} \to \infty$ can be found in the literature \cite{trice_prb07,Seric_pof2018}; in Appendix~\ref{thicksub_sect:convergence} we demonstrate the convergence of our numerical results to this analytical solution as $H_{\rm s}$ increases.

Figures~\ref{thicksub_fig:flat_film_contours}(c) and (d) show peak average film temperatures and LL for the substrate whose thermal conductivity varies with temperature according to Eq.~\eqref{thicksub_TC_expression}. The trend of peak temperature and LL is similar to the $\kappa=1$ results shown in Figs.~\ref{thicksub_fig:flat_film_contours}(a) and (b), although the temperatures are much lower and thus the LL is shortened for given (${\rm Bi},H_{\rm s})$ pairs. For the entire simulation $\kappa(T_{\rm s})\geq 1$, so that substrate diffusion occurs more rapidly, and heat is then transferred faster away from the film, compared with the $\kappa=1$ case. This becomes increasingly important when considering films that evolve, since viscosity may depend strongly on temperature \cite{allaire_jfm2021}.
Finally, it should be noted that some temperatures in Fig.~\ref{thicksub_fig:flat_film_contours} surpass the boiling point of the film ($T_{\rm boil}\approx 2.088$), while our model neglects possible evaporation. Although models that account for evaporation exist (see, e.g.~\cite{oron_rmp97} for a review), in practice the laser fluence is often adjusted to the system of interest so that no significant mass is lost to evaporation. These results, therefore, can serve as a guideline for such fluence adjustments.

\subsection{2D Evolving films}\label{thicksub_sect:2d_evolving_films}
In this section the film surface is initially prescribed by Eq.~\eqref{thicksub_h_ic1}, with $\delta=0.01$, on the spatial domain $x\in [-\pi,\pi]$, and we investigate the influence of ${\rm Bi}$ and $H_{\rm s}$ on the film evolution. The initially solid film is static until it melts, at which point it evolves according to Eq.~\eqref{thicksub_thin_film}. Once the film re-solidifies, its evolution 
stops. To maintain generality, we allow the material parameters governing surface tension, viscosity and thermal conductivity to vary with average film temperature, so that $\Gamma=\Gamma(t)$  via Eq.~\eqref{thicksub_surf_tens_dep} and $\mathcal{M}=\mathcal{M}(t)$ via Eq.~\eqref{thicksub_viscosity_eq}. Similarly, the thermal conductivity of the substrate is allowed to depend on substrate temperature, $\kappa=\kappa(T_{\rm s})$ (see Eq.~\eqref{thicksub_TC_expression} in Appendix~\ref{thicksub_sect:thermal_conduct} for the form used).
\begin{figure}[thb!]
    \centering
    \includegraphics[width=0.8\textwidth]{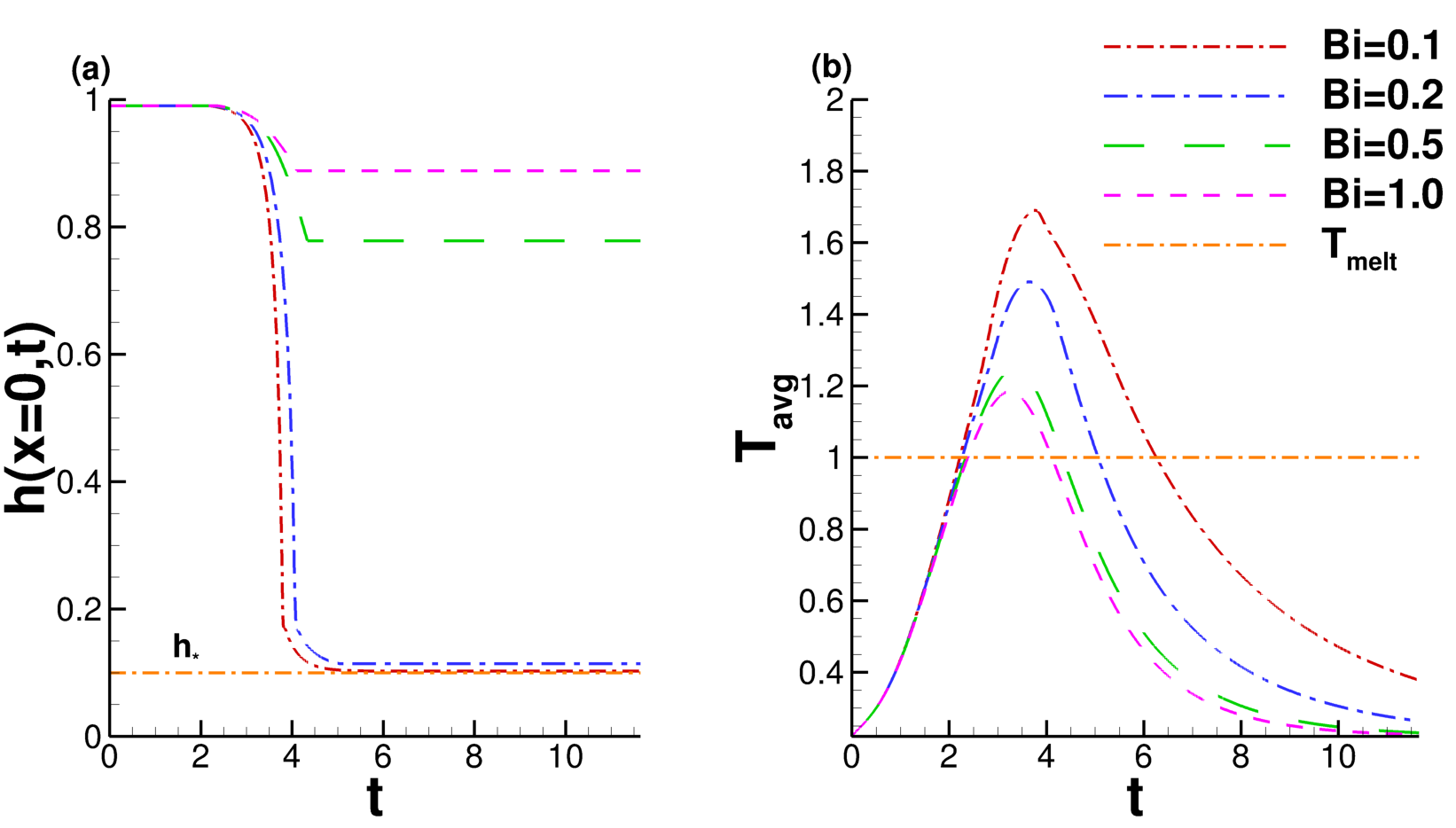}
    \caption{(a) Evolution of film thickness at $x=0$ for ${\rm Bi}=0.1$ (black), $0.2$ ({\color{red} red}, dash-dotted), $0.5$ ({\color{blue} blue} dash-dotted), $1.0$ ({\color{green} green} dashed); and equilibrium film thickness $h=h_*$ ({\color{orange} orange} dot-dashed). (b) Average film temperature corresponding to the cases shown in (a). The material parameters are variable, $\Gamma=\Gamma(t), \mathcal{M}=\mathcal{M}(t), \kappa=\kappa(T_{\rm s})$, substrate thickness is fixed, $H_{\rm s}=10$, and melting temperature, $T_{\rm melt}=1$ ({\color{orange} orange} dot-dashed).}
    \label{thicksub_fig:T_avg_and_h_evolving_films_Bi}
\end{figure}

Figures \ref{thicksub_fig:T_avg_and_h_evolving_films_Bi}(a) and (b) show the evolution of the film midpoint $(x=0)$ and the average film temperature, respectively, for various values of ${\rm Bi}$ and for fixed substrate thickness, $H_{\rm s}=10$. The trend of shorter LL in Fig.~\ref{thicksub_fig:T_avg_and_h_evolving_films_Bi} as ${\rm Bi}$ 
increases is consistent with Fig.~\ref{thicksub_fig:flat_film_contours}(d). Consequently, the films for ${\rm Bi}=0.5$ and ${\rm Bi}=1.0$ solidify prior to any significant evolution, whereas for ${\rm Bi}=0.1$ the film dewets fully. For ${\rm Bi}=0.2$ the film mostly dewets, but solidifies just before its surface reaches the equilibrium film thickness, $h=h_*$. This intricate balance between solidification and dewetting highlights the importance of the value of ${\rm Bi}$ in determining whether full or partial dewetting occurs.
\begin{figure}[thb!]
    \centering
    \includegraphics[width=0.8\textwidth]{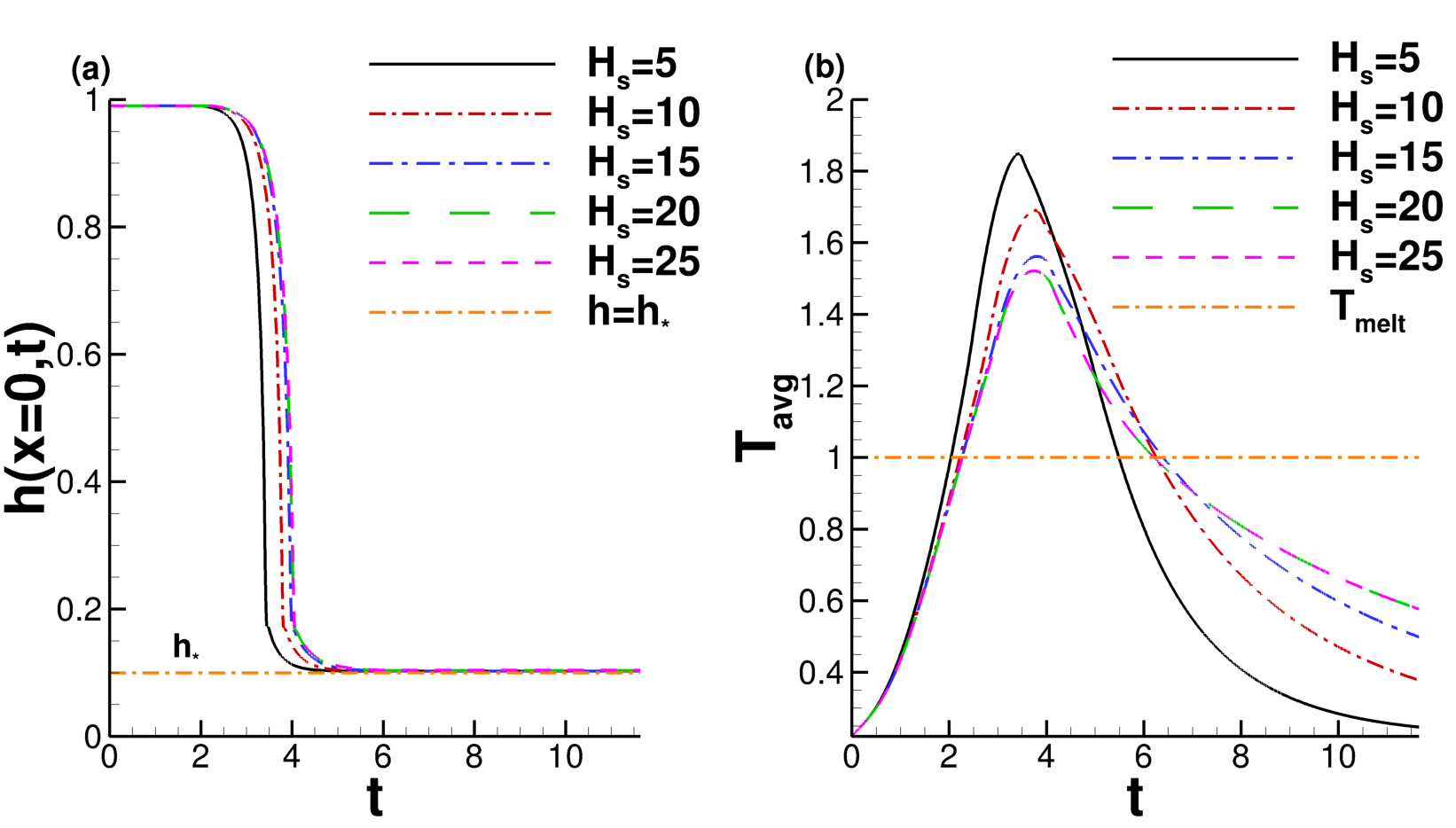}
    \caption{(a) Evolution of film thickness at $x=0$ for $H_{\rm s}=5$ (black), $10$ ({\color{red} red}, dash-dotted), $15$ ({\color{blue} blue} dash-dotted), $20$ ({\color{green} green} dashed), $25$ ({\color{magenta} magenta} dashed), and the equilibrium film thickness $h=h_*$ ({\color{orange} orange} dot-dashed). (b) Average film temperature corresponding to the $H_{\rm s}$ cases in (a) and melting temperature, $T_{\rm melt}$ ({\color{orange} orange} dot-dashed). The material parameters are variable, $\Gamma=\Gamma(t), \mathcal{M}=\mathcal{M}(t), \kappa=\kappa(T_{\rm s})$, and ${\rm Bi}=0.1$.}
    \label{thicksub_fig:T_avg_and_h_evolving_films_Hs}
\end{figure}

Next, we consider the influence of substrate thickness. Similarly to
Fig.~\ref{thicksub_fig:T_avg_and_h_evolving_films_Bi}, Figs.~\ref{thicksub_fig:T_avg_and_h_evolving_films_Hs}(a) and (b) show the midpoint film thickness and average film temperature, but for varying $H_{\rm s}$. Here the Biot number is fixed at ${\rm Bi}=0.1$. From Fig.~\ref{thicksub_fig:T_avg_and_h_evolving_films_Hs}(a), we see that increasing substrate thickness increases the dewetting speed by only a small amount. Since in Fig.~\ref{thicksub_fig:T_avg_and_h_evolving_films_Hs}(b) films on thinner substrates are seen to achieve higher temperatures, the film on the thinnest substrate, $H_{\rm s}=5$, has lowest viscosity and dewets fastest in (a). The observed increase in peak temperature with substrate thickness, and the similar LLs for ${\rm Bi}=0.1$, are consistent with Figs.~\ref{thicksub_fig:flat_film_contours}(c) and (d). For completeness, we include the analog of Fig.~\ref{thicksub_fig:T_avg_and_h_evolving_films_Hs} for the case ${\rm Bi}=0.2$ in Appendix~\ref{thicksub_Bi_0.2_case} and show that the findings are again consistent.

To summarize, varying substrate thickness ($H_{\rm s}$) and heat loss from the lower surface (Bi) may result in films that solidify prior to complete dewetting. We will see in Section~\ref{thicksub_3D_evolving_films} that the substrate thickness 
may play a significant role in determining the final configurations of the 3D films.

\subsection{3D Evolving films}\label{thicksub_3D_evolving_films}
\begin{figure}[thb!]
    \centering
    \includegraphics[width=0.6\textwidth]{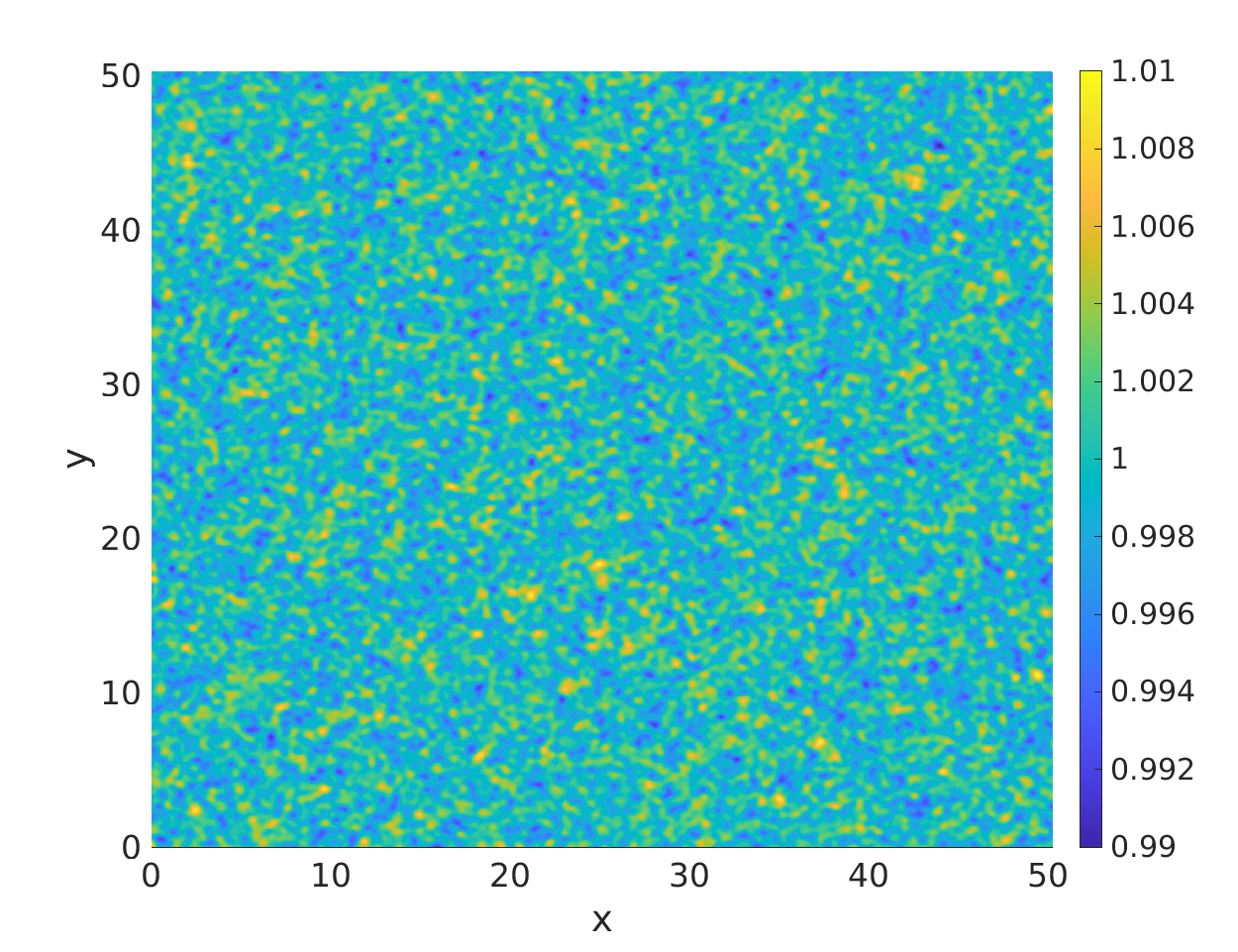}
    \caption{Initial film thickness $h(x,y,0)$ for 3D simulations, described by random noise perturbations to the flat film $h=1$, and given by Eq.~\eqref{thicksub_random_ic_eq}. The domain has length $16\pi$ in both $x$ and $y$ directions.}
    \label{thicksub_fig:random_ic}
\end{figure}

Next, we consider the role of the temperature-dependent material parameters, the substrate thickness, $H_{\rm s}$, and the Biot number, Bi, in the pattern formation for 3D films, with free surface $z=h(x,y,t)$. For this section, we consider randomly perturbed films with the initial free surface disturbance specified by Eq.~\eqref{thicksub_random_ic_eq} (shown in Fig.~\ref{thicksub_fig:random_ic}), and follow the same melting/solidification procedure described in Section~\ref{thicksub_sect:2d_evolving_films}. In all cases, the domain is a square of linear dimension $16 \pi$, surface tension is a function of average film temperature via Eq.~\eqref{thicksub_surf_tens_dep} and, except where otherwise specified, the Biot number is fixed at ${\rm Bi}=0.1$. We consider both constant viscosity $\mathcal{M}=1$ and (average) temperature-dependent viscosity $\mathcal{M}(t)$ (see Eq.~\eqref{thicksub_viscosity_eq}), and $\kappa=1, \kappa(T_{\rm s})$ for substrate thermal conductivity.

In earlier work~\cite{Seric_pof2018,allaire_jfm2021}, 2D simulations reveal that temperature-dependent viscosity is crucial for modeling the correct dewetting speed of the films. We now confirm the importance of accounting for temperature-dependent viscosity in 3D simulations. 

\subsubsection{Influence of viscosity}
\begin{figure}[thb!]
    \centering
    \includegraphics[width=\textwidth]{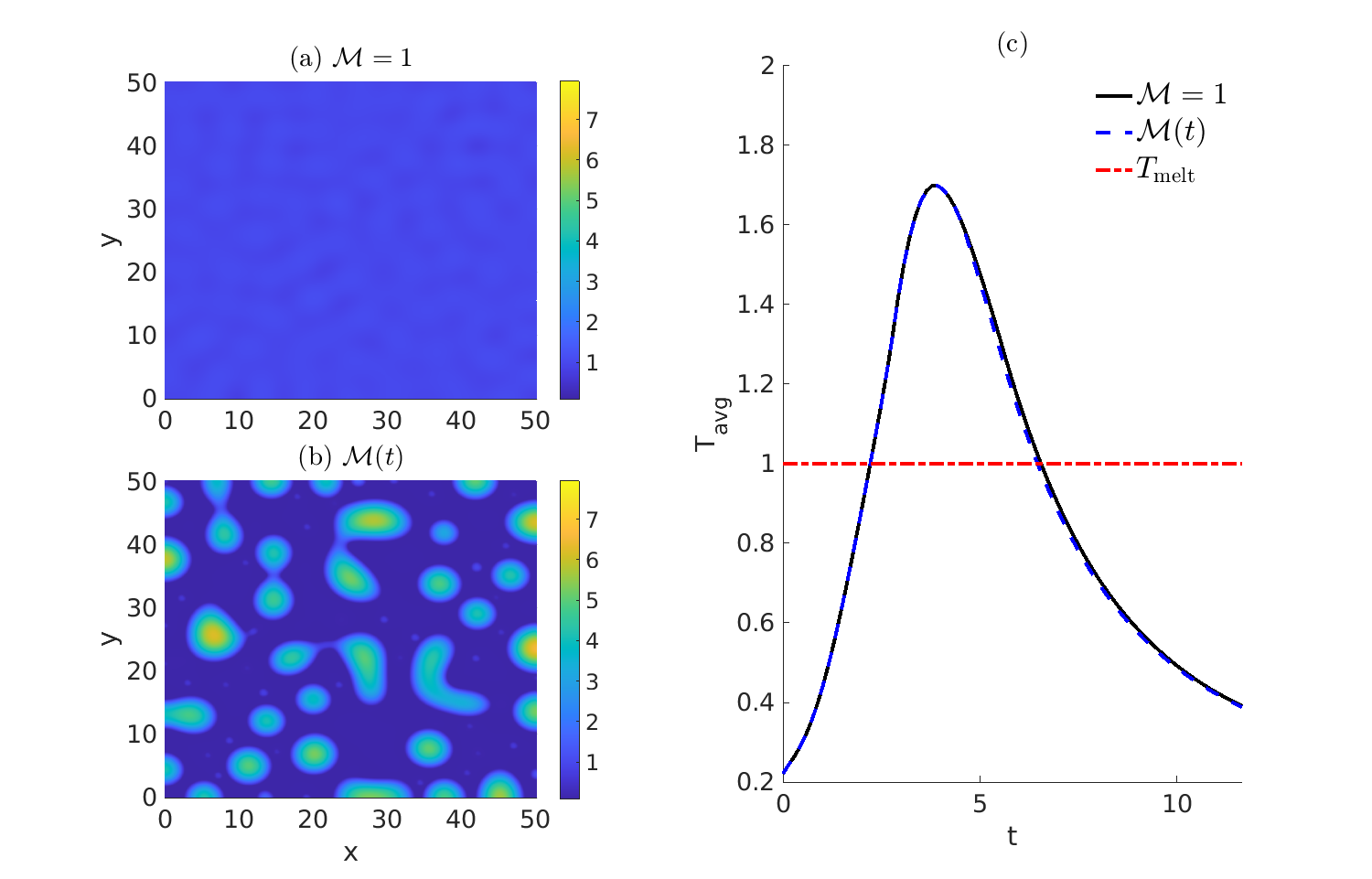}
    \caption{Final film thickness for (a) $\mathcal{M}=1$ and (b) $\mathcal{M}(t)$. Average film temperatures are shown in (c), with melting temperature, $T_{\rm melt}$. In (a) the film solidifies before significant evolution occurs, whereas in (b) further dewetting occurs with some droplet formation. Both films are initialized with the same random noise (Eq.~\eqref{thicksub_random_ic_eq}, shown in Fig.~\ref{thicksub_fig:random_ic}). The LLs are approximately $4.35$ and $4.15$ for (a) and (b), respectively, and in both cases ${\rm Bi}=0.1$ and $H_{\rm s}=10$. For animations of the film evolution, see Supplementary materials, movie3.
        }
    \label{thicksub_fig:gpu_viscosity}
\end{figure}

Figures~\ref{thicksub_fig:gpu_viscosity}(a) and (b) both show the final solidified film for $\kappa=\kappa(T_{\rm s})$ but (a) corresponds to $\mathcal{M}=1$ (viscosity fixed at melting value) and (b) to $\mathcal{M}=\mathcal{M}(t)$ (viscosity depends on average temperature as given by Eq.~\eqref{thicksub_viscosity_eq}). The main finding is that the variable-viscosity film in Fig.~\ref{thicksub_fig:gpu_viscosity}(b) has mostly dewetted and formed droplets prior to resolidification, whereas the constant-viscosity film in Fig.~\ref{thicksub_fig:gpu_viscosity}(a) has barely evolved. Figure \ref{thicksub_fig:gpu_viscosity}(c) shows the average film temperature $T_{\rm avg}$ in both cases, along with the melting temperature, $T_{\rm melt}$; we see that $T_{\rm avg}$ is nearly identical for the two cases, despite the very different fluid dynamics. Since the final film structures are very different but the LLs are nearly identical, we conclude that the variable viscosity is crucial for accurate modeling of dewetting within the liquid phase. Note that the spatially-varying form of viscosity, $\mathcal{M}(x,t)$, given by Allaire {\it et al.}~\cite{allaire_jfm2021}, which replaces $T_{\rm avg}$ by $T_{\rm f}$ in Eq.~\eqref{thicksub_viscosity_eq}, produces essentially identical results to Fig.~\ref{thicksub_fig:gpu_viscosity}(b), due to the weak in-plane spatial variation of film temperature (result shown in Appendix~\ref{thicksub_sect:spat_varying_visc} for completeness).

\subsubsection{Influence of thermal conductivity}
\begin{figure}[thb!]
    \centering
    \includegraphics[width=\textwidth]{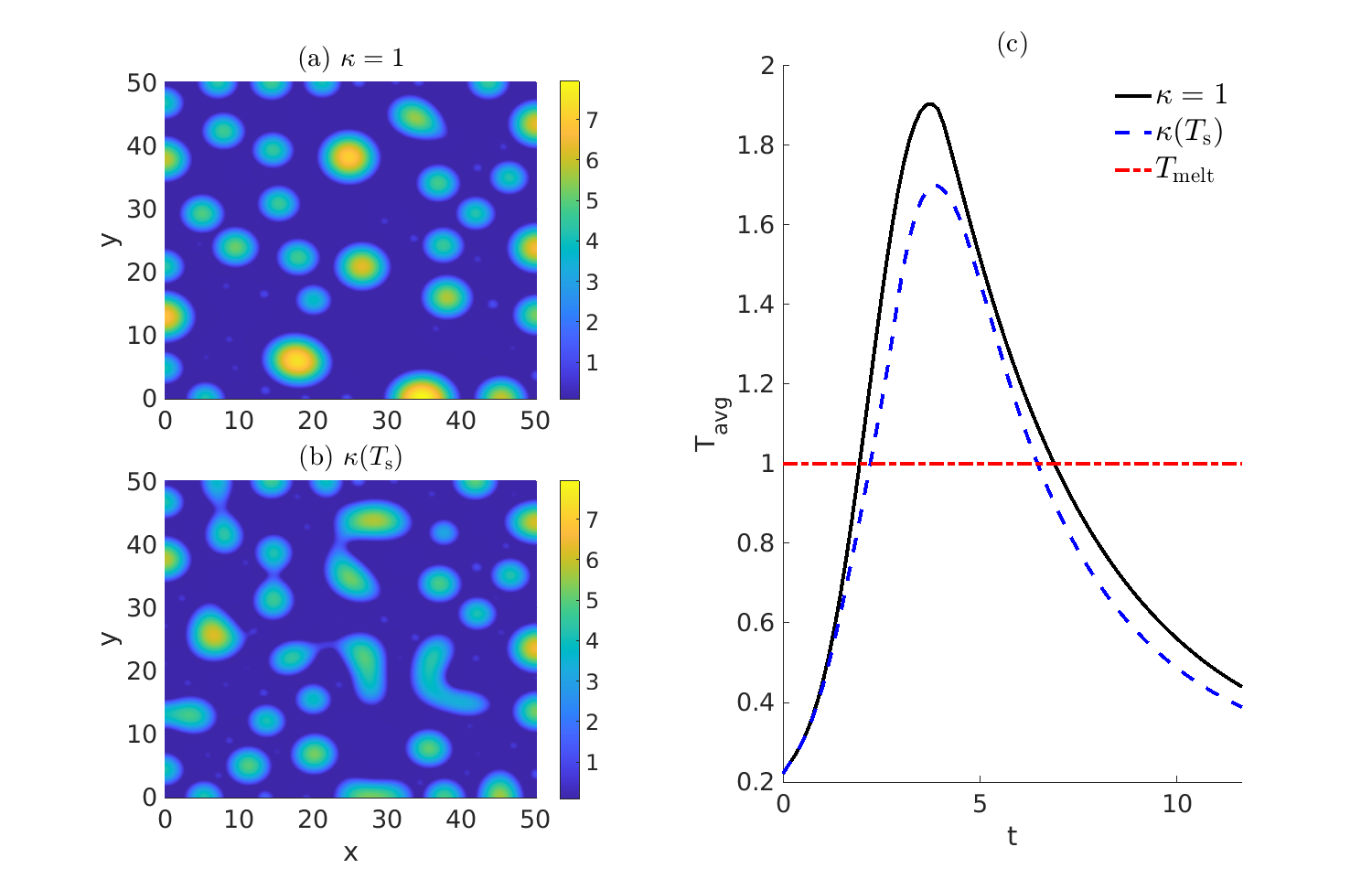}
    \caption{Final film thickness for (a) $\kappa=1$ and (b) $\kappa(T_{\rm s})$. Average film temperatures are shown in (c), with melting temperature, $T_{\rm melt}$. Here, $\mathcal{M}=\mathcal{M}(t)$, (b) is identical to Fig.~\ref{thicksub_fig:gpu_viscosity}(b) and the LL for (a) is approximately $4.82$. In both cases ${\rm Bi}=0.1$ and $H_{\rm s}=10$. For animations, see Supplementary materials, movie4.
    }
    \label{thicksub_fig:influence_of_TC}
\end{figure}

Next, we consider the influence of temperature-dependent substrate thermal conductivity on film dewetting behavior. Figure \ref{thicksub_fig:influence_of_TC} shows final solidified film thickness for (a) constant, and (b) temperature-varying ($\kappa(T_{\rm s})$), substrate thermal conductivity, each with temperature-dependent viscosity $\mathcal{M}=\mathcal{M}(t)$. Figure \ref{thicksub_fig:influence_of_TC}(c) shows the average film temperature over time for both cases. The decreased LL and lower peak temperature for $\kappa(T_{\rm s})$ are consistent with the flat film results in Figs.~\ref{thicksub_fig:flat_film_contours}(c) and (d), although the difference is not dramatic. Despite this, dewetting has clearly proceeded further in (a) than in (b), as evidenced by the differences in film heights: dewetting in case (b) is slower due to the higher film viscosity resulting from lower temperatures. Coarsening is also more advanced in case (a) at solidification, with generally larger droplets than case (b), due to both premature solidification in case (b) and to different values of the surface tension parameter $\Gamma$, known to alter instability wavelengths \cite{allaire_jfm2021}. 

\subsubsection{Influence of substrate thickness}
\begin{figure}[thb!]
    \centering
    \includegraphics[width=\textwidth]{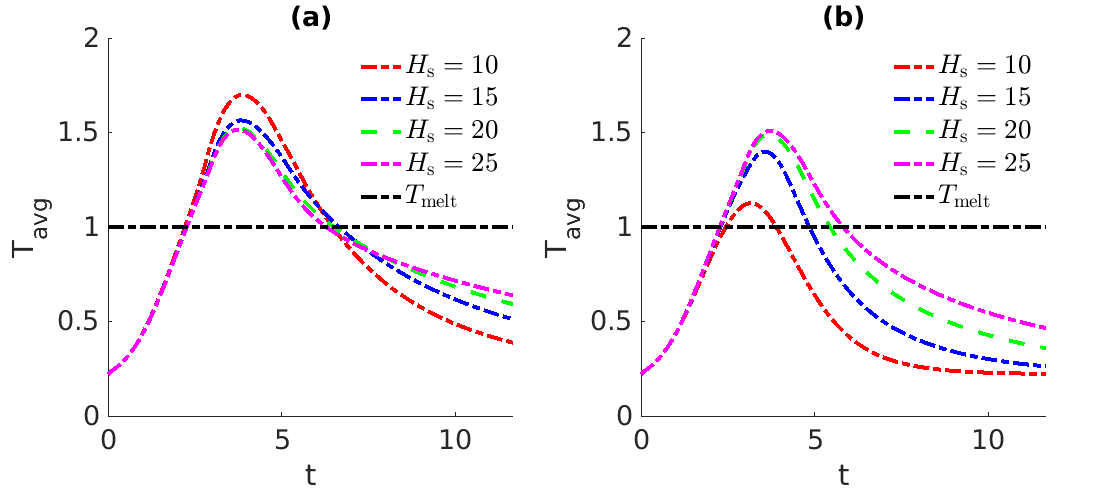}
    \caption{Average film temperatures, $T_{\rm avg}$, for (a) ${\rm Bi}=0.1$ and (b) ${\rm Bi}=10^{3}$, when deposited on substrates of thickness $H_{\rm s}=10$ ({\color{red} red} dot dashed line), $H_{\rm s}=15$ ({\color{blue} blue} dot dashed line), $H_{\rm s}=20$ ({\color{green} green} dashed line), and $H_{\rm s}=25$ ({\color{violet} magenta} dot dashed line). The melting temperature is given by the black dot dashed line.}
    \label{thicksub_fig:influence_of_sub_Tavg}
\end{figure}

\begin{figure}[thb!]
    \centering
    \includegraphics[width=\textwidth]{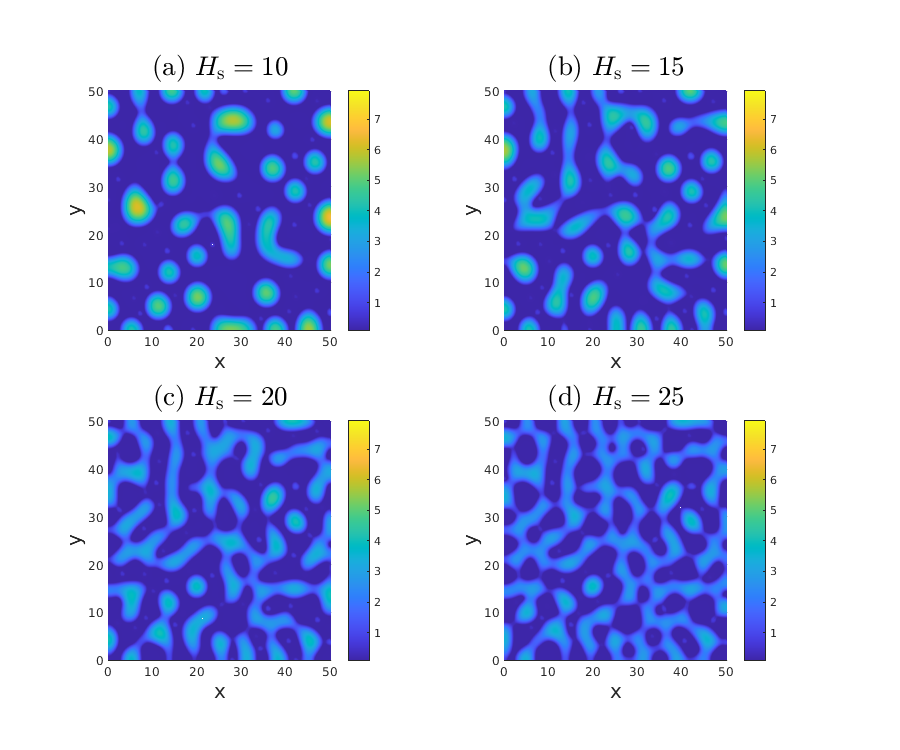}
    \caption{Final film thickness for ${\rm Bi}=0.1$, and on substrates of thickness (a) $H_{\rm s}=10$, (b) $H_{\rm s}=15$, (c) $H_{\rm s}=20$, and (d) $H_{\rm s}=25$, with temperature-dependent material parameters, $\Gamma(t)$, $\mathcal{M}(t)$, and $\kappa(T_{\rm s})$. Films on thicker substrates dewet slower due to the lower temperatures (and higher viscosity), see Fig.~\ref{thicksub_fig:influence_of_sub_Tavg}(a). Here, (a) is the same as Fig.~\ref{thicksub_fig:gpu_viscosity}(b) and the LLs are (a) $t=4.15$, (b) $t=4.31$, (c) $t=4.19$, and (d) $t=3.95$.  For animations, see Supplementary materials, movie5.}
    \label{thicksub_fig:sub_thickness_sweep}
\end{figure}

\begin{figure}[thb!]
    \centering
    \includegraphics[width=\textwidth]{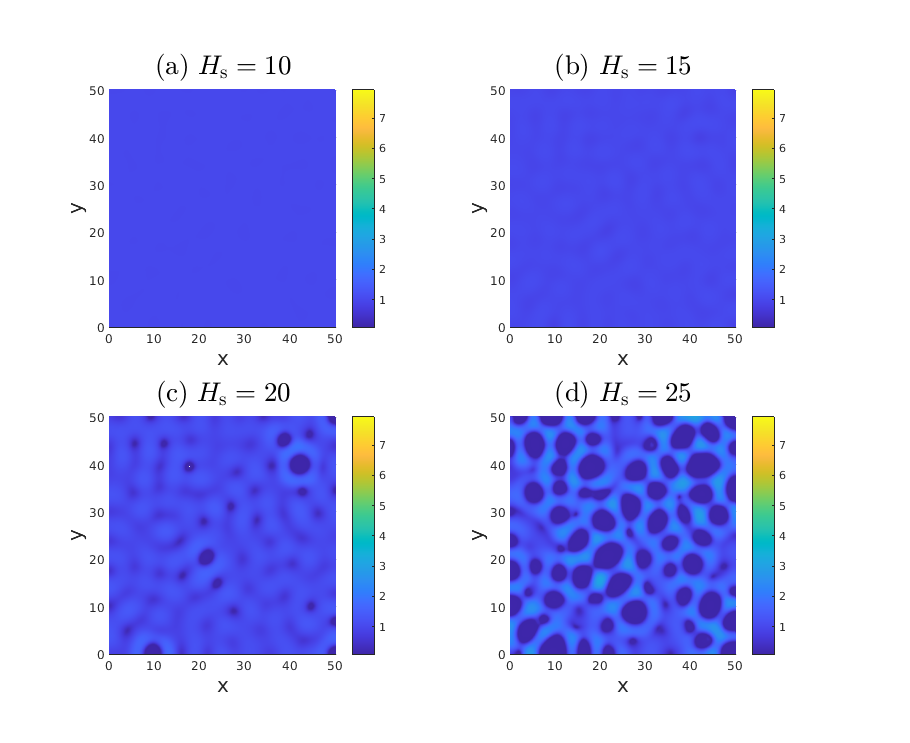}
    \caption{Final film thickness on poorly insulated substrates, ${\rm Bi}=10^{3}$, of thickness (a) $H_{\rm s}=10$, (b) $H_{\rm s}=15$, (c) $H_{\rm s}=20$, and (d) $H_{\rm s}=25$, with temperature-dependent material parameters, $\Gamma(t)$, $\mathcal{M}(t)$, and $\kappa(T_{\rm s})$.  Films on thicker substrates dewet faster due to the higher temperatures (and lower viscosity), see Fig.~\ref{thicksub_fig:influence_of_sub_Tavg}(b). The LLs are (a) $t=1.46$, (b) $t=2.57$, (c) $t=3.16$, and (d) $t=3.52$. For animations, see Supplementary materials, movie6.}
    \label{thicksub_fig:influence_of_sub_thickness_Bi1e3}
\end{figure}

Figures~\ref{thicksub_fig:influence_of_sub_Tavg}, \ref{thicksub_fig:sub_thickness_sweep}, and \ref{thicksub_fig:influence_of_sub_thickness_Bi1e3} illustrate the role of $H_{\rm s}$ on the dewetting process for small (${\rm Bi}=0.1$) and large (${\rm Bi}=10^{3}$) values of the Biot number. Figure \ref{thicksub_fig:influence_of_sub_Tavg}(a) shows average film temperatures for a well-insulated substrate, ${\rm Bi}=0.1$, and $H_{\rm s}=10$, 15, 20, and $25$, where both film viscosity and substrate thermal conductivity are temperature-dependent, $\mathcal{M}=\mathcal{M}(t)$ and $\kappa=\kappa(T_{\rm s})$. The similar LLs and small variations in peak temperature observed are nearly identical to those for the 2D film in Fig.~\ref{thicksub_fig:T_avg_and_h_evolving_films_Hs}(b). Nevertheless, the small deviations in peak temperature as $H_{\rm s}$ varies are important because of the strong temperature dependence of viscosity, which changes the dewetting speed.

Figure~\ref{thicksub_fig:influence_of_sub_Tavg}(b) similarly shows average film temperature for the same substrate thicknesses as in (a) but for a poorly-insulated substrate, ${\rm Bi}=10^{3}$. The significantly decreased temperatures and shorter LLs for thinner substrates are consistent with Fig.~\ref{thicksub_fig:flat_film_contours}(c). Note in particular the reversal of the trend between Figs.~\ref{thicksub_fig:influence_of_sub_Tavg}(a) and (b), with peak temperature decreasing with $H_{\rm s}$ in (a), and increasing with $H_{\rm s}$ in (b). In Fig.~\ref{thicksub_fig:influence_of_sub_Tavg}(b), the peak temperatures are generally lower and the LLs much shorter, which (we now show) may lead to different final solidified film configurations.

Figure~\ref{thicksub_fig:sub_thickness_sweep} shows the final solid film configurations for (a) $H_{\rm s}=10$, (b) $H_{\rm s}=15$, (c) $H_{\rm s}=20$ and (d) $H_{\rm s}=25$, for ${\rm Bi}=0.1$ (corresponding to Fig.~\ref{thicksub_fig:influence_of_sub_Tavg}(a)). Since average peak temperature $T_{\rm peak}$ decreases with $H_{\rm s}$, the dewetting speed decreases from (a)--(d) due to the viscosity increase. This is to some degree surprising,  since the influence of $H_{\rm s}$ was not readily apparent in the 2D case. The proposed explanation is that, in our 3D simulations, we prescribe a random initial condition, and therefore it takes time for the fastest growing mode of instability to develop. This surplus time slows the dewetting sufficiently for the thicker substrates that it is still incomplete at resolidification.

Figure~\ref{thicksub_fig:influence_of_sub_thickness_Bi1e3} shows the final solid film configurations for (a) $H_{\rm s}=10$, (b) $H_{\rm s}=15$, (c) $H_{\rm s}=20$ and (d) $H_{\rm s}=25$ for the poorly insulating substrate, ${\rm Bi}=10^{3}$, corresponding to Fig.~\ref{thicksub_fig:influence_of_sub_Tavg}(b). Since $T_{\rm peak}$ now increases with substrate thickness, viscosity decreases and dewetting speed increases from (a)-(d). In this case, none of the simulations (a)-(d) fully dewet (recall the lower peak temperatures in Fig.~\ref{thicksub_fig:influence_of_sub_Tavg}(b) compared with Fig.~\ref{thicksub_fig:influence_of_sub_Tavg}(a) leading to earlier resolidification in Fig.~\ref{thicksub_fig:influence_of_sub_thickness_Bi1e3} compared with Fig.~\ref{thicksub_fig:sub_thickness_sweep}). The films in Figs.~\ref{thicksub_fig:influence_of_sub_thickness_Bi1e3}(c) and (d) begin to form holes, but those in (a) and (b) barely evolve.  Collectively,
Figs.~\ref{thicksub_fig:sub_thickness_sweep} and~\ref{thicksub_fig:influence_of_sub_thickness_Bi1e3} indicate that the final configuration of the resolidified film depends on both $H_{\rm s}$ and ${\rm Bi}$ in a nontrivial way.

\section{Conclusions}\label{thicksub_sect:conclusion}
We have modeled and simulated the evolution of pulsed laser irradiated nanoscale metal films that are deposited on thick substrates. In particular, we have focused on the role that the underlying substrate plays in determining both the temperature of the film and its corresponding evolution. With regards to material parameters, our model accounts for temperature dependence of both surface tension and viscosity of the film. Our 3D simulations indicate that if temperature dependence of viscosity is not included, the films may not fully dewet.

The film liquid lifetime (LL) and spatially-averaged peak temperature ($T_{\rm peak}$) are found to depend on the substrate heat loss (as characterized by a Biot number, Bi, governing heat loss at the lower surface), substrate thickness $H_{\rm s}$, and the thermal conductivity model used (specifically, whether it is taken to be constant, or varying with temperature). $T_{\rm peak}$ is found to vary strongly with Bi, but less so with $H_{\rm s}$. In particular, we find that the correlation between $H_{\rm s}$ and $T_{\rm peak}$ changes from negative to positive according to whether the substrate is well-insulated (Bi~$\ll 1$) or poorly-insulated (Bi~$\gg 1$). The choice of well- or poorly-insulated substrates can lead to significantly different final solidified film configurations. Including temperature-varying thermal conductivity, in general, increases the heat loss from the film to the substrate, decreasing $T_{\rm peak}$ and therefore liquid lifetimes. The decreased film temperatures observed with temperature-varying thermal conductivity lead to a much smaller film viscosity, which reduces the speed of dewetting. Our 3D simulations show that this can lead to films that solidify prematurely, although the effect is not as dramatic as that of changing $H_{\rm s}$. Interestingly, we find that varying $H_{\rm s}$ does not appear to alter significantly the LL of the films; however, a small but significant change in $T_{\rm peak}$ results, which again alters viscosity and thus the final configuration of the film.

Our model omits a number of effects, the possible relevance of which we briefly discuss. First, we neglected temperature-dependent thermal conductivity of the metal film. Although this could be added to the model, with notable added complexity to the numerical schemes described in Appendices~\ref{thicksub_sect:2d_numerics} and~\ref{thicksub_sect:3d_numerics}, the modest changes to thermal conductivity \cite{Powell_1966} would be inconsequential on the fast time scale of heat transfer across the film. Second, our simulations assume that phase change occurs instantaneously. In practice, partial melting and solidification may occur, in different parts of the film. The current model could be altered to include such effects, most readily by modifying the form of Eq.~\eqref{thicksub_viscosity_eq} to account for spatial variations in film temperature, and viscosities that increase dramatically when the film temperature drops below $T_{\rm melt}$. Radiative heat losses and evaporation are also neglected in the modeling; both effects may become important for certain choices of film materials. Finally, in-plane diffusion is neglected in the substrate. These additional effects should be considered in future work.

\section*{Acknowledgement}
This research was supported by NSF DMS-1815613 (L.J.C., L.K., R.H.A.); by NSF CBET-1604351 (R.H.A., L.K.), and by CNMS2020-A-00110 (L.K.). This research was, in part, conducted at the Center for Nanophase Materials Sciences, which is a DOE Office of Science User Facility. Computational resources used a Director Discretionary allocation with the Summit supercomputer at Oak Ridge National Laboratory.

\section*{Declaration of Interests}
The authors report no conflict of interest.

\section*{Author ORCID}
R.H. Allaire, https://orcid.org/0000-0002-9336-3593, L.J. Cummings, https://orcid.org/0000-0002-7783-2126, and L. Kondic, https://orcid.org/0000-0001-6966-9851.
\appendix
%\section{Appendix}

%====================================%
%== Thick Substrate Appendices=======%
%====================================%

\section{Values of parameters}
\label{app_table}
    %=======================================================================%
%--Table of Parameters used for Model Comparison-------------------------%
%=======================================================================%
\begin{table}[H]
\centering
{\renewcommand{\arraystretch}{0.75} % for the vertical padding
  \begin{tabular}{ | l | l | l | l |} \hline 
 \textbf{Parameter} & \textbf{Notation} & \textbf{Value} & \textbf{Unit}  \\ \hline 
  Viscosity at Melting Temperature & $\mu_{\rm f}$ \cite{Dong_prf16} & $4.3 \times 10^{-3}$ & $\mathrm{Pa\cdot s}$  \\  
  Surface tension at Melting Temperature & $\gamma_{\rm f}$ \cite{Dong_prf16} & 1.303 & $\mathrm{J\cdot m^{-2}}$ \\ 
  Wavelength of maximum growth & $\lambda_{\rm m}=2\pi/ \sqrt{\Omega \epsilon h_*^2/(2Lh_0^4) \left( 2 h_0 - 3  \right)}$ \cite{allaire_jfm2021} & $180.84$ & $\mathrm{nm}$  \\ 
  Vertical length scale & $H$ & $10$ & $\mathrm{nm}$  \\  
  Horizontal length scale & $L=\lambda_{\rm m}/(2\pi)$ & $28.78$ & $\mathrm{nm}$  \\  
  Time scale & $t_{\rm s} = 3L\mu_{\rm f}/ (\epsilon^3 \gamma_{\rm f})$ & $6.79$ & $\mathrm{ns}$ \\  
   Temperature scale/Melting Temperature & $T_{\rm melt}$ & $1358$ & $\mathrm{K}$ \\ 
    
  Film density & $\rho_{\rm f}$ \cite{Dong_prf16} & $8000$ & $\mathrm{kg \cdot m^{-3}}$  \\  
  SiO$_2$ density & $\rho_{\rm s}$ \cite{Dong_prf16}  & $2200$ & $\mathrm{kg \cdot m^{-3}}$  \\ 
  Film specific heat capacity & $c_{\rm f}$ \cite{Dong_prf16} & $495$ & $\mathrm{J \cdot kg^{-1} \cdot K^{-1}}$ \\  
  SiO$_2$ specific heat capacity & $c_{\rm s}$ \cite{Dong_prf16} & $937$ & $\mathrm{J \cdot kg^{-1} \cdot K^{-1}}$ \\  
  Film heat conductivity & $\kappa_{\rm f}$ \cite{Dong_prf16} & $340$ & $\mathrm{W \cdot m^{-1} \cdot K^{-1}}$  \\ 
    SiO$_2$ heat conductivity & $\kappa_{\rm s}$ \cite{Dong_prf16} & $1.4$ & $\mathrm{W \cdot m^{-1} \cdot K^{-1}}$\\  
  Film absorption length & $\alpha_{\rm f}^{-1} H$ \cite{Dong_prf16} & $11.09$ & $\mathrm{nm}$  \\ 
  Temp. Coeff. of Surf. Tens. & $ \gamma_{\rm T}$ \cite{Dong_prf16} & $-0.23 \times 10^{-3}$ & $\mathrm{J \cdot m^{-2} \cdot K^{-1}}$  \\  
    Hamaker constant & $A_{\rm H}$ \cite{lang13} & $3.49\times 10^{-17}$ & $\mathrm{J}$  \\ 
  Reflective coefficient & $r_0$ \cite{Dong_prf16} & $0.3655$ & 1  \\ 
  Film reflective length & $\alpha_{\rm r}^{-1} H$ \cite{Dong_prf16} & $12.0$ & $\mathrm{nm}$ \\  
    Laser energy density & $E_0$ \cite{lang13} & $1400$ & $\mathrm{J \cdot m^{-2}}$  \\ 
    Gaussian pulse peak time & $t_{\rm p} t_{\rm s}$ \cite{lang13} & $18$ & $\mathrm{ns}$ \\ 
    Equilibrium film thickness & $h_* H$ & $1$ & $\mathrm{nm}$  \\  
  Mean film thickness & $h_0 H$ & $10$ & $\mathrm{nm}$ \\  
  SiO$_2$ thickness & $H_{\rm s} H$ & $50-250$ & $\mathrm{nm}$  \\ 
  Room temperature & $T_{\rm a} T_{\rm melt}$ & $300$ & $\mathrm{K}$ \\  
  SiO$_2$ Heat Transfer Coefficient & $\alpha_{\rm s}$ & $10^{5}-10^{11}$ & $\mathrm{W \cdot m^{-2} \cdot K^{-1}}$ \\ 
    Characteristic Velocity & $U$ & $4.237$ & $\mathrm{m \cdot s^{-1}}$ \\ 
    Activation Energy & $E$ & $30.5$ & ${\rm kJ} \cdot {\rm mol}^{-1}$  \\ \hline
 \end{tabular} }
 \caption{Parameters used for liquid Cu film and SiO$_2$ substrate.}
 \label{thicksub_table:ref_paras}
\end{table}

\section{Model validity: neglecting in-plane heat diffusion in the substrate}
\label{thicksub_sect:model_validation}

For brevity, we denote the asymptotically-reduced model described by Eqs.~\eqref{thicksub_asymptotic model1}-\eqref{thicksub_asymptotic_model_end} as model (A). In our previous work on this system \cite{allaire_jfm2021}, it was assumed that the film is placed upon a substrate sufficiently thin that neglecting in-plane diffusion in the substrate is asymptotically valid. In Section~\ref{thicksub_sect:modeling} of the present work, we allow the underlying substrate to be thick relative to the film, so the neglect of terms representing in-plane diffusion in the substrate requires further justification. For this purpose,  we consider a model, denoted (FA) (here, ``F'' indicates that a ``full'' (2D) model is used for heat flow in the substrate, while ``A'' denotes the ``asymptotically'' reduced model that applies to heat transport in the film), which includes Eq.~\eqref{thicksub_asymptotic model1} and Eqs.~\eqref{thicksub_cont_temp_nd}--\eqref{thicksub_asymptotic_model_end}, but replaces Eq.~\eqref{thicksub_asymptotic_sub_eqn} with a full 2D heat transport model in the substrate,
\begin{align}
    {\rm Pe}_{\rm s}\partial_t T_{\rm s} = \epsilon^2 \partial_x^2 T_{\rm s} + \partial_z^2 T_{\rm s}. \label{thicksub_model_FA}
\end{align}

\begin{figure}[thb!]
    \centering
    \includegraphics[width=0.8\textwidth]{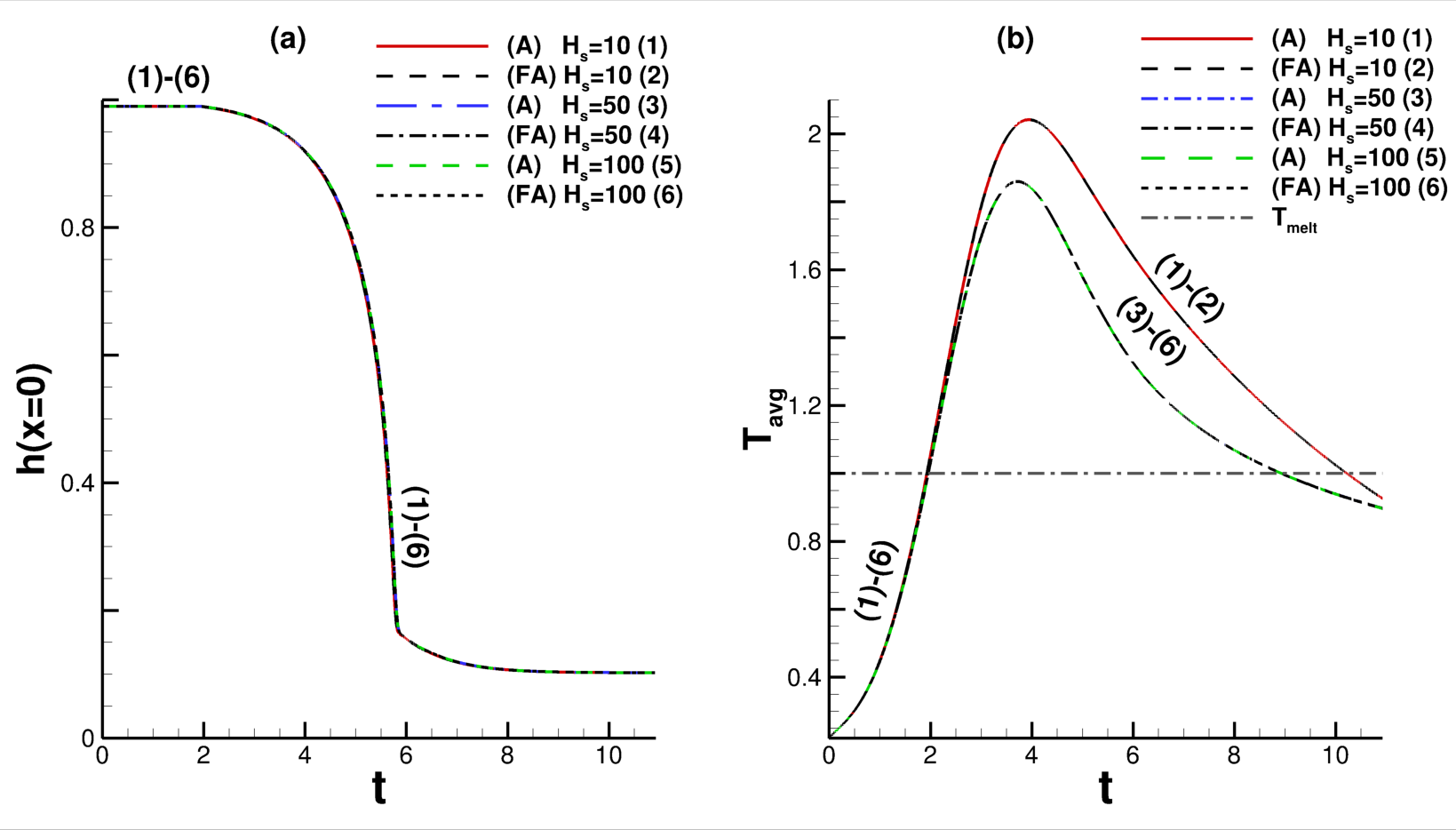}
    \caption{(a) Film thickness at the midpoint, $x=0$, and (b) average film temperature for model (A) (Eqs.~\eqref{thicksub_asymptotic model1}--\eqref{thicksub_asymptotic_model_end}) and model (FA) (where Eq.~\eqref{thicksub_asymptotic_sub_eqn} is replaced by Eq.~\eqref{thicksub_model_FA}). Models (A) and (FA) agree for substrate thicknesses $H_{\rm s}=10, 50,$ and $100$. For both (a) and (b) the parameters were held constant, $\Gamma=\mathcal{M}=\kappa=1$, and ${\rm Bi}\approx 4.3\times 10^{-2}$. Note that the temperature range here differs from that in other plots.}
    \label{thicksub_fig:appendix_model_comparison}
\end{figure}

 Figure~\ref{thicksub_fig:appendix_model_comparison} shows the evolution of the film thickness at the midpoint, $x=0$, (a) and average film temperature (b) for 2D films on substrates of thicknesses $H_{\rm s}=10,50,$ and $100$. In (a), the film is initially given by Eq.~\eqref{thicksub_h_ic1} and $h(0,t)$ is determined by solving Eq.~\eqref{thicksub_thin_film} with $\Gamma=\mathcal{M}=\kappa=1$ and ${\rm Bi}\approx 4.3 \times 10^{-2}$. The heat conduction is solved using both models (A) and (FA). We find good agreement between model (A) (the thermal model used in the main text) and model (FA). This indicates that including lateral diffusion in the substrate does not influence the film (neither evolution nor heating) and can be neglected. 

\begin{figure}[thb!]
    \centering
    \includegraphics[width=\textwidth]{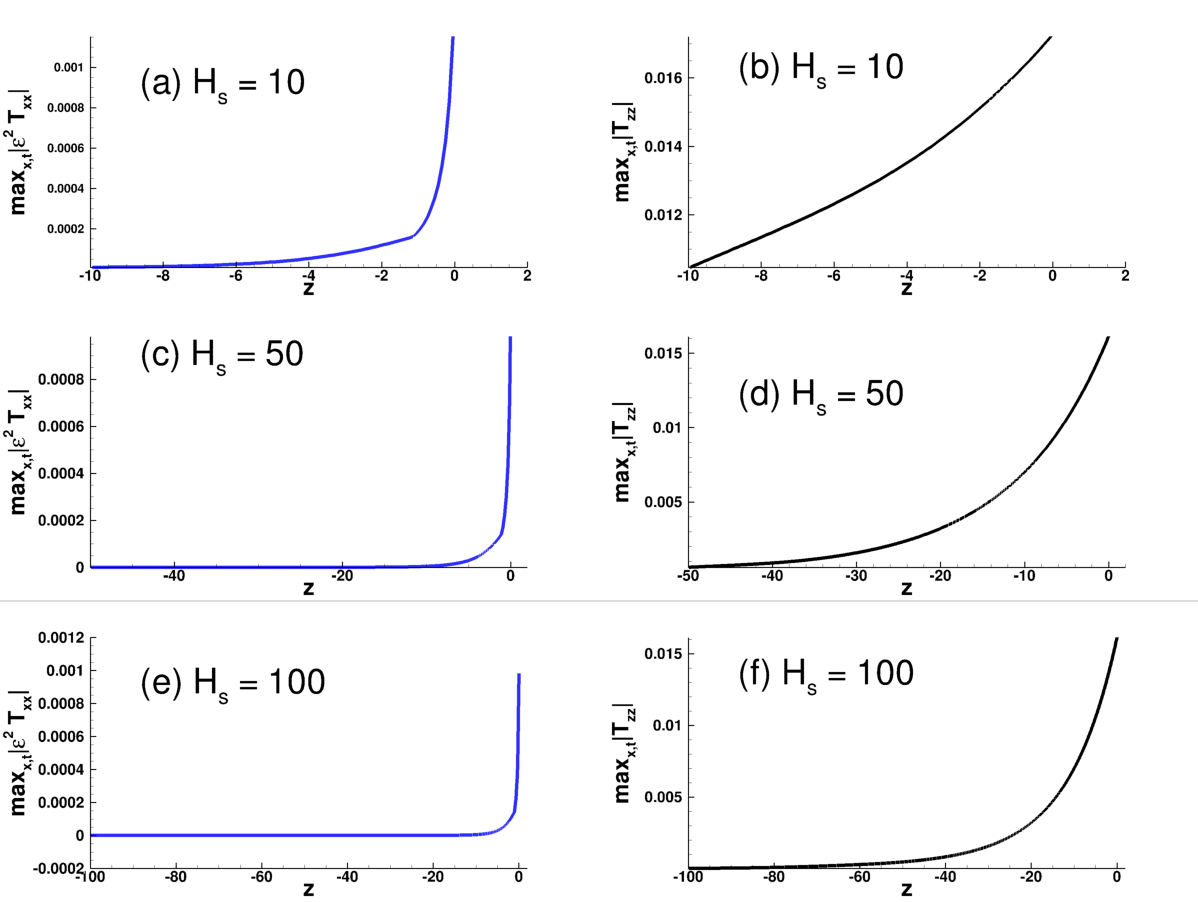}
    \caption{Maximum magnitude of in-plane diffusion over all $x$ and $t$, $\mbox{max}_{x,t}|\epsilon^2 T_{xx}|$, for (a) $H_{\rm s}=10$, (b) $H_{\rm s}=50$, (c) $H_{\rm s}=100$; and maximum out-of-plane diffusion, $\mbox{max}_{x,t}|T_{zz}|$ similarly for (b), (d), and (f). Out-of-plane diffusion is orders of magnitude larger than in-plane diffusion. The parameters are the same as Fig.~\ref{thicksub_fig:appendix_model_comparison}.}
    \label{thicksub_fig:Txx_vs_Tzz}
\end{figure}

To further justify dropping lateral diffusion in the substrate, we simulate full 2D heat conduction in both the film and substrate in the same case given in Fig.~\ref{thicksub_fig:appendix_model_comparison}. Figure \ref{thicksub_fig:Txx_vs_Tzz} shows the largest value in magnitude of both in-plane diffusion, $\max_{x,t}|\epsilon^2 T_{xx}|$ ({\color{blue} blue}), and out-of-plane diffusion, $\max_{x,t}|T_{zz}|$ (black), as a function of $z$, for $H_{\rm s}=10$ (a,~b), $H_{\rm s}=50$ (c,~d), and $H_{\rm s}=100$ (e,~f). In all cases, the term representing in-plane diffusion in the substrate is at least $10$ times smaller than that representing out-of-plane diffusion. The former, then, can be dropped without significant loss of accuracy. 

\section{Temperature-varying thermal conductivity}\label{thicksub_sect:thermal_conduct}
The dimensionless substrate thermal conductivity, given by $\kappa(T_{\rm s})$, depends on the local values of the substrate temperature $T_{\rm s}$. Limited data exist on ${\rm SiO}_2$ thermal conductivity values at high temperatures (e.g. higher than film melting temperature) and the wide range of temperatures observed during film heating presents a modeling challenge. To determine the appropriate functional dependence for $\kappa(T_{\rm s})$ we follow the approach of Combis {\it et al}~\cite{Combis2012}, which utilizes both the annealing temperature, $T_{\rm anneal}$, and the softening temperature, $T_{\rm soften}$. The values we use are based on changes in the thermal expansion coefficient \cite{Combis2012}, although in practice these temperatures are measured by a sudden change in various material properties (such as viscosity), which could occur in such a wide range of temperatures considered. For more general information regarding $T_{\rm anneal}$ and $T_{\rm soften}$ see
e.g.~Callister~\cite{Callister_2007} or Petrie~\cite{Petrie2007}. Based on the data provided by the manufacturer (Silica Suprasil 312 Type 2~\cite{Heraeus_2019}), we use $T_{\rm anneal}=1.03$ and $T_{\rm soften}=1.40$, respectively (all temperatures are normalized by the film melting temperature used in our simulations and thermal conductivity is normalized by the value at melting temperature, $\kappa_{\rm s}$).
\begin{figure}[thb!]
    \centering
    \includegraphics[width=0.8\textwidth]{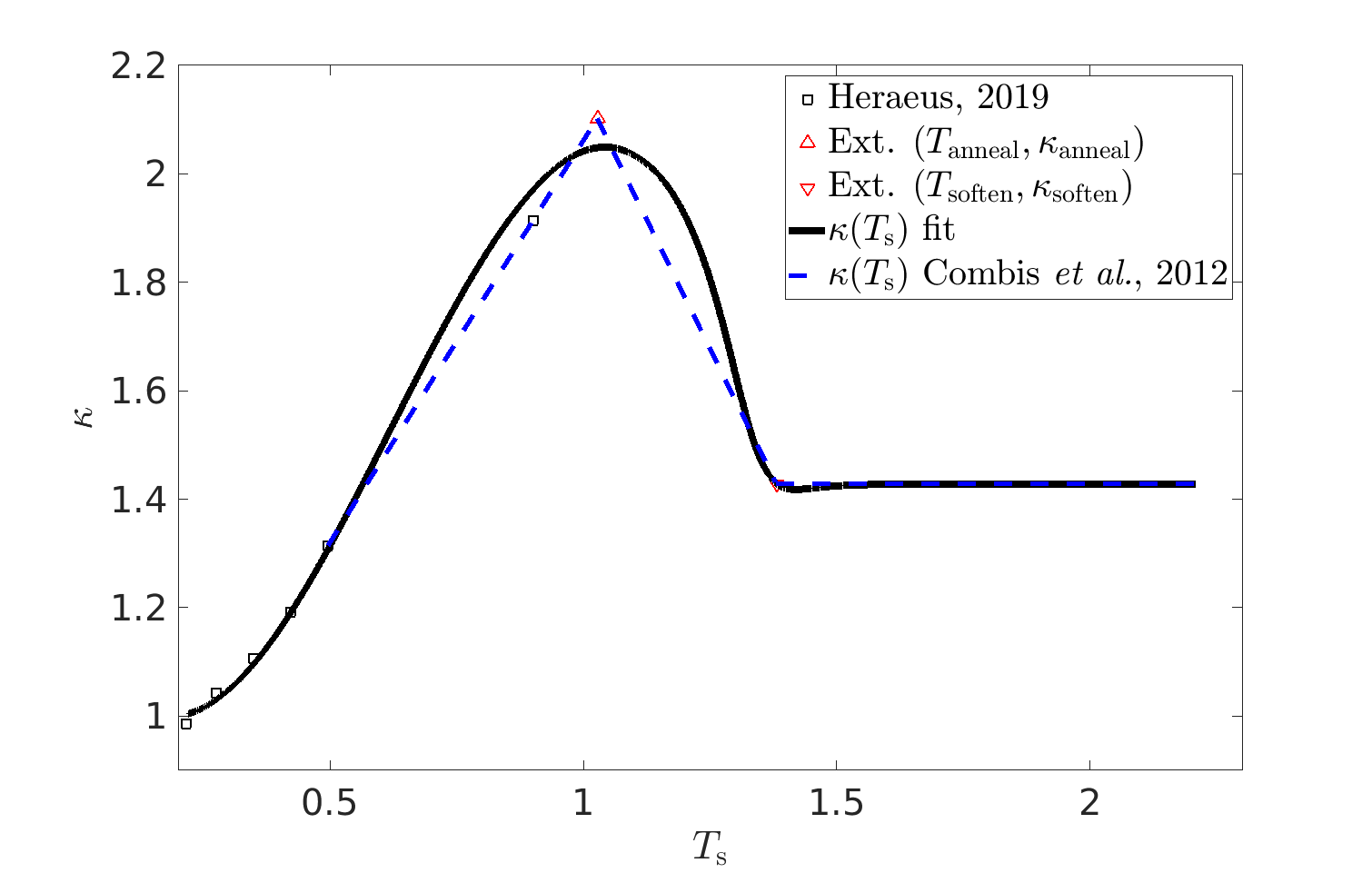}
    \caption{Manufacturer data of thermal conductivity at various temperatures (black $\square$), extrapolated values at annealing temperature ({\color{red} red $\bigtriangleup$}) and softening temperature ({\color{red} red $\bigtriangledown$}), the fit of substrate thermal conductivity to temperature $T_{\rm s}$ used in this manuscript (black solid line), and the fit used by Combis {\it et al.}~\cite{Combis2012} ({\color{blue} blue} dashed line).  Note that $T_{\rm s}$ is in units of $T_{\rm melt}$, so that the leftmost point on the horizontal axis corresponds to the ambient temperature, where $\kappa =1$). 
        }
    \label{thicksub_fig:TC_data}
\end{figure}
Figure \ref{thicksub_fig:TC_data} shows the data provided (black squares) by the manufacturer~\cite{Heraeus_2019}, the piecewise linear fit used by Combis {\it et al.}~\cite{Combis2012} (blue dashes), and the form of $\kappa(T_{\rm s})$ we use (black solid line). Instead of using a piecewise linear profile, we use a cubic polynomial smoothed with sigmoid functions, in the following form:
\begin{align}
    \kappa(T_{\rm s}) = \frac{1}{1 + \exp\left(\beta_1 T_{\rm s}-\beta_2 \right)}(a + b T_{\rm s} + c T_{\rm s}^2 + d T_{\rm s}^3) + \frac{1}{1 + \exp\left(\beta_2-\beta_1 T_{\rm s}\right)}\kappa_{\rm soften},\label{thicksub_TC_expression}
\end{align}
where $a,b,c,d,\beta_2$ are fitting parameters, $\beta_1$ is a scaling factor, and $\kappa_{\rm soften}$ is the thermal conductivity at softening temperature, all of which are given in Table~\ref{thicksub_table:TC_paras_table}. This form captures the thermal conductivity at low, annealing, and softening temperatures reasonably well and provides a large range of values for use in simulations. Note that above the softening temperature the thermal conductivity is nearly constant, a simplifying assumption made due to lack of reliable data in this regime.

\begin{table}[H]
\centering
  \begin{tabular}{ | l | l | l | l |} \hline 
\textbf{Parameter} & \textbf{Notation} & \textbf{Value} \\ \hline 
    Fitting Parameter & $a$ & $-1.23\times 10^{-4}$ \\ \hline 
    Fitting Parameter & $b$ & $2.06\times 10^{-1}$\\ \hline
    Fitting Parameter & $c$ & $-59.89$ \\ \hline
    Fitting Parameter & $d$ & $3.22 \times 10^{4}$ \\ \hline
    Scaling Factor & $\beta_1$ & $30.12$ \\ \hline
    Fitting Parameter & $\beta_2$ & $40.0$ \\ \hline
    SiO$_2$ Thermal conductivity at $T_{\rm soften}$ & $\kappa_{\rm soften}$ & $1.43$ \\ \hline
    SiO$_2$ Annealing Temperature & $T_{\rm anneal}$ & $1.03$ \\ \hline 
    SiO$_2$ Softening temperature & $T_{\rm soften}$ & $1.40$ \\ \hline
 \end{tabular}
 \caption{Table of parameters used for the fit of temperature-dependent thermal conductivity, given by Eq.~\eqref{thicksub_TC_expression}.} \label{thicksub_table:TC_paras_table}
\end{table} 

\section{Relevance of radiative losses}\label{thicksub_sect:radiative_losses}
Here we briefly consider the relevance of radiative heat losses at the film surface, $z=h$. For simplicity we consider a simple energy argument. Consider the case of a flat film $h=1$, which is at melting temperature. The total internal thermal energy of the system is then $\rho_{\rm f} c_{\rm f} T_{\rm melt} L^2 H$. The rate of energy loss at the boundary $z=h$ due to radiation is proportional to the fourth power of temperature and is given by $\sigma_{\rm SB} \varepsilon_r T_{\rm melt}^4 \left( 1 - T_{\rm a}^4 \right) L^2$, where $\sigma_{\rm SB}=5.67\times 10^{-8} {\rm W m^{-2} K^{-4}}$ is the Stefan-Boltzmann constant and $\varepsilon_r \approx 0.14$ is the thermal emissivity \cite{metals_ref_book_2004}. In time interval $\Delta t$ then, the ratio of the energy lost to free surface radiation and the internal thermal energy is,
\begin{align}
    r_{\rm rad} = \frac{\Delta t \sigma_{\rm SB} \varepsilon_r T_{\rm melt}^4 \left( 1 - T_{\rm a}^4 \right)}{H \rho_{\rm f} c_{\rm f} T_{\rm melt}}.
\end{align}
For the parameter values in our problem, the timescale $\Delta t$ on which these two energies become comparable, $r_{\rm rad}=O(1)$, is found to be $\Delta t \approx 2\times 10^{-3} {\rm s}$, a millisecond time interval, which is 
five orders of magnitude longer than the laser pulse and dewetting time scales of interest in this work. 
Therefore, radiative losses can be safely neglected.

\section{Convergence results}\label{thicksub_sect:convergence}
\begin{figure}[thb!]
    \centering
    \includegraphics[width=0.8\textwidth]{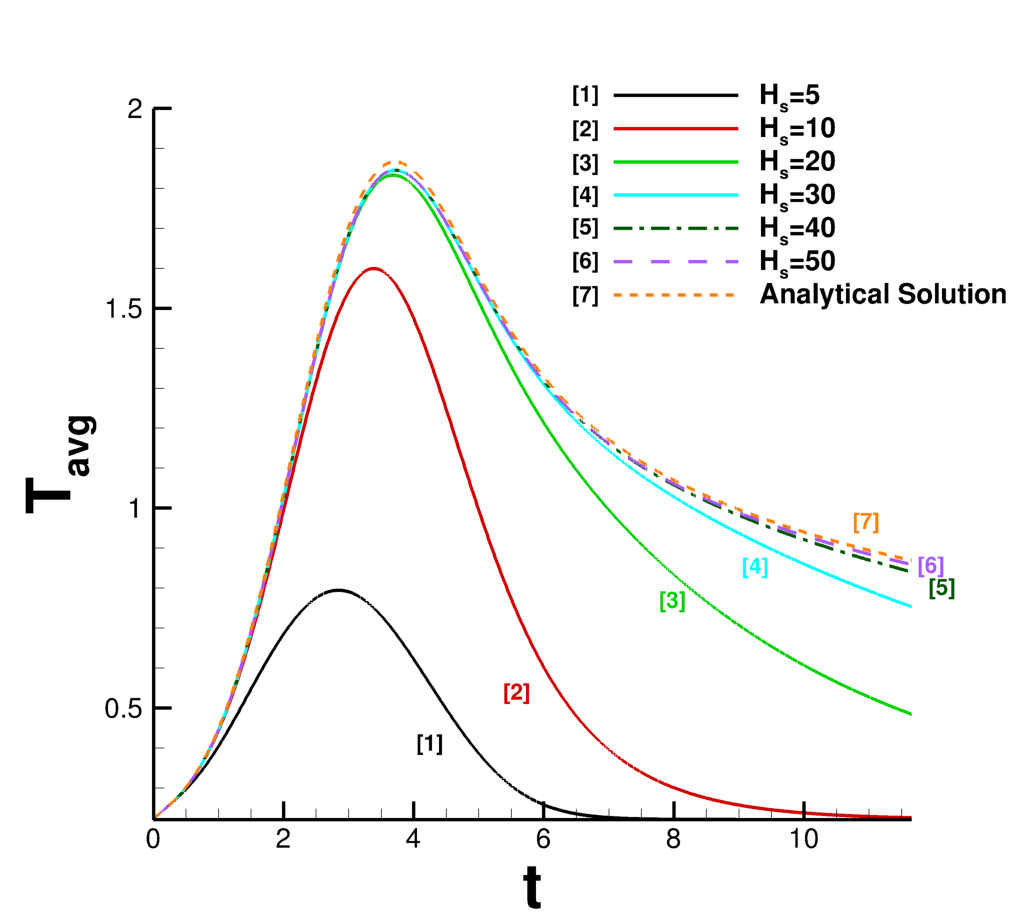}
    \caption{Average film temperature, $T_{\rm avg}$, for varying substrate thickness; and the analytical 
    solution, see text, in the limit $H_{\rm s}, {\rm Bi} \to \infty$. In each simulation, ${\rm Bi}=10^{8}$.}
    \label{thicksub_fig:analytical_convergence}
\end{figure}

Here we show that $T_{\rm avg}$ from our model converges to the analytical solution~\cite{trice_prb07,Seric_pof2018} in the limit $H_{\rm s},{\rm Bi} \to \infty$ and for a uniform flat film, $h=h_0$. Figure \ref{thicksub_fig:analytical_convergence} plots average film temperature for $H_{\rm s}=5,10,20,30,40,50$ as well as the analytical solution. As substrate thickness is increased, the average film temperatures converge to the analytical result, as expected.

\section{Influence of substrate thickness for ${\rm Bi}=0.2$}\label{thicksub_Bi_0.2_case}
\begin{figure}[hb!]
    \centering
    \includegraphics[width=0.8\textwidth]{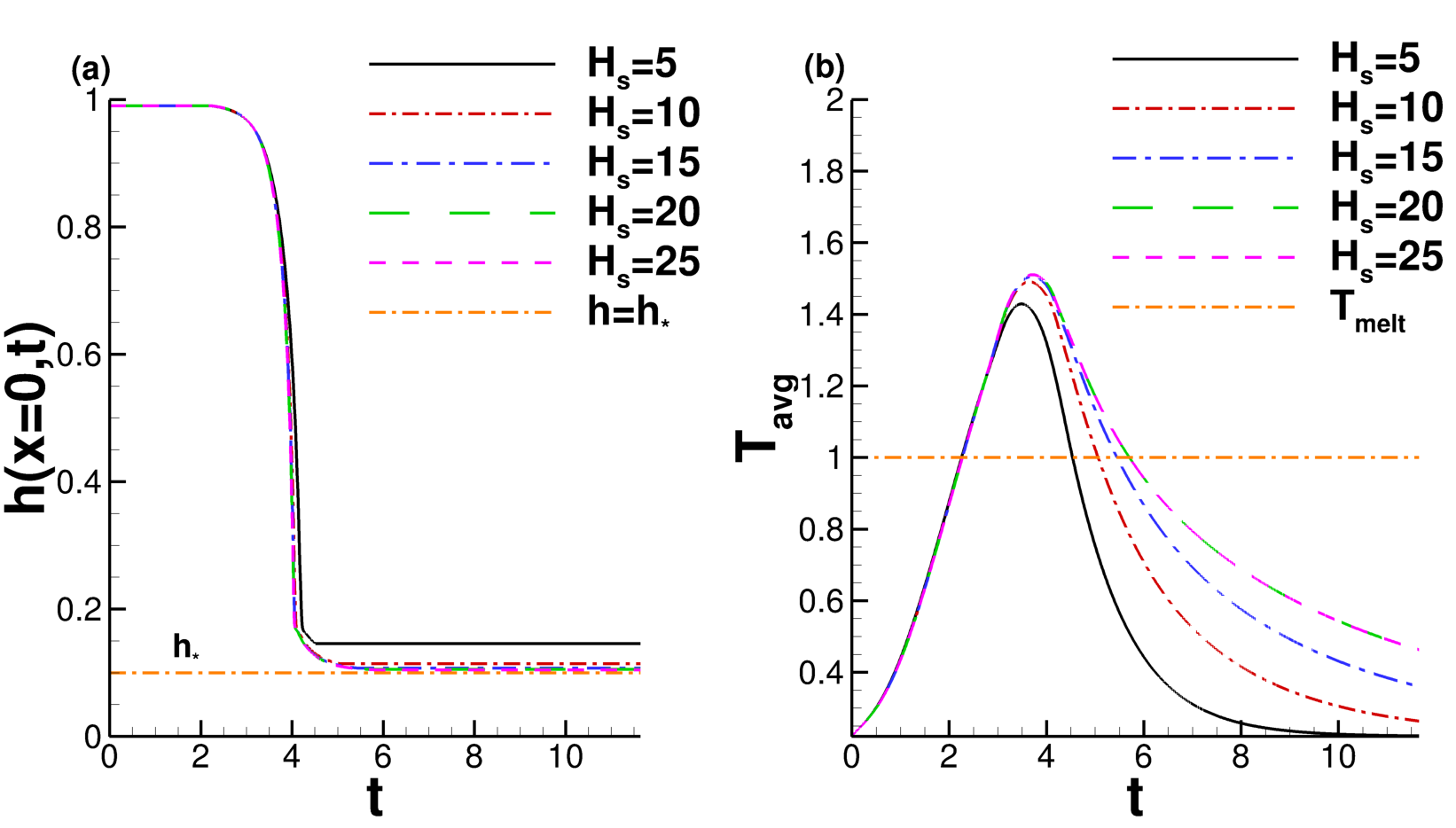}
    \caption{(a) Evolution of film thickness at $x=0$ for $H_{\rm s}=5$ (black), $H_{\rm s}=10$ ({\color{red} red}, dash-dotted), $H_{\rm s}=15$ ({\color{blue} blue} dash-dotted), $H_{\rm s}=20$ ({\color{green} green} dashed), $H_{\rm s}=25$ ({\color{magenta} magenta} dashed), and $h=h_*$ ({\color{orange} orange} dot-dashed). (b) Average film temperature corresponding to the $H_{\rm s}$ cases in (a) and melting temperature, $T_{\rm melt}$ ({\color{orange} orange} dot-dashed). The material parameters are variable, $\Gamma=\Gamma(t), \mathcal{M}=\mathcal{M}(t), \kappa=\kappa(T_{\rm s})$, and ${\rm Bi}=0.2$.}\label{thicksub_fig:T_avg_and_h_evolving_films_Hs_Bi_1e-2}
\end{figure}

Figures~\ref{thicksub_fig:T_avg_and_h_evolving_films_Hs_Bi_1e-2}(a) and (b) show the evolution of the film midpoint and average temperatures, respectively, for five different substrate thicknesses $H_{\rm s}=5,10,15,20,25$ as in Fig.~\ref{thicksub_fig:T_avg_and_h_evolving_films_Hs}, but now for ${\rm Bi}=0.2$. We see in Fig.~\ref{thicksub_fig:T_avg_and_h_evolving_films_Hs_Bi_1e-2}(b) that the liquid lifetimes vary more significantly than for ${\rm Bi}=0.1$, but the effect of varying $H_{\rm s}$ is still small relative to that of varying ${\rm Bi}$ (compare Fig.~\ref{thicksub_fig:T_avg_and_h_evolving_films_Bi}(a)). Of the $H_{\rm s}$ cases considered, the film for $H_{\rm s}=5$ shows the largest difference (similar to Fig.~\ref{thicksub_fig:T_avg_and_h_evolving_films_Hs}(a)). In contrast to the ${\rm Bi}=0.1$ case, here the film with $H_{\rm s}=5$ solidifies before full dewetting ($h(0,t)$ does not reach the equilibrium film thickness). Finally, note that despite the weak influence of $H_s$ on film evolution, a small change in LL may signal premature solidification of the film,  as we see in 3D simulations (e.g.~Fig.~\ref{thicksub_fig:influence_of_sub_Tavg}).

\section{Influence of spatially varying viscosity in 3D}\label{thicksub_sect:spat_varying_visc}

\begin{figure}[thb!]
    \centering
    \includegraphics[width=\textwidth]{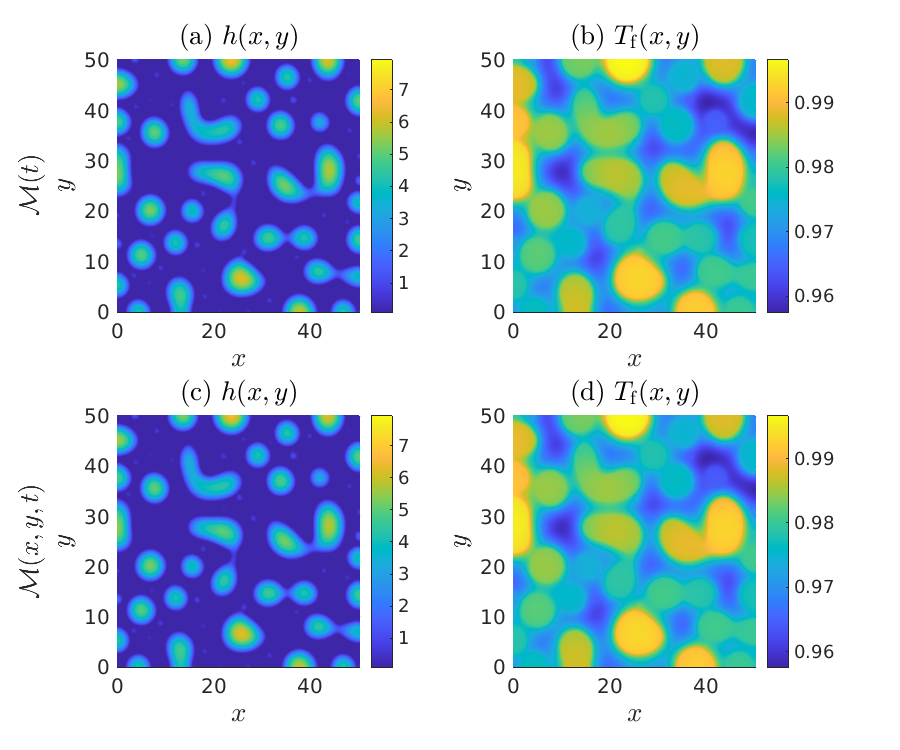}
    \caption{Final film thickness and film temperature, $T_{\rm f}$ for $\mathcal{M}(t)$ (a) and (b) and $\mathcal{M}(x,y,t)$ (c) and (d). $T_{\rm f}$ is plotted at the time of solidification. Here $\Gamma=1$, thermal conductivity is variable $\kappa=\kappa(T_{\rm s})$, and ${\rm Bi}=0.1$.}
    \label{thicksub_fig:spat_varying_visc_plot}
\end{figure}

Here we briefly consider the effect of spatially varying viscosity, where $T_{\rm avg}(t)$ is replaced by $T_{\rm f}(x,y,t)$ in the viscosity law, Eq.~\eqref{thicksub_viscosity_eq}. Figures~\ref{thicksub_fig:spat_varying_visc_plot}(a) and (b) show film thickness and film temperature 
at the final solidification time in the case where viscosity depends only on average film temperature, $\mathcal{M}=\mathcal{M}(t)$  (Fig.~\ref{thicksub_fig:spat_varying_visc_plot}(a) is identical to Fig.~\ref{thicksub_fig:gpu_viscosity}(b)). Figures~\ref{thicksub_fig:spat_varying_visc_plot}(c) and (d) show the corresponding film thickness and temperature for the spatially-varying viscosity case, $\mathcal{M}(x,y,t)$. There is no noticeable difference between the film thicknesses in (a) and (c), nor between the temperatures in (b) and (d). Note that the spatial variation of temperature is small in (b) and (d). Consequently, $T_{\rm avg} (t)$ is a good approximation of $T_{\rm f}(x,y,t)$ in Eq.~\eqref{thicksub_viscosity_eq}.

\section{Numerical schemes and initial condition implementation}
\label{app:numerics}

\subsection{2D Numerical schemes including temperature-dependent thermal conductivity}\label{thicksub_sect:2d_numerics}

Here, we describe the numerical schemes used to solve for the film height, $h$, temperature, $T_{\rm f}$, and substrate temperature, $T_{\rm s}$. First, we describe the spatial discretization, and then the solution mechanism for $T_{\rm s}$ and $T_{\rm f}$. We conclude with the numerical scheme used to compute $h$.
For notational simplicity, we drop the arguments $(x,y,z,t)$ remembering that the dependent 
variables are space- and time-dependent. 

\begin{figure}[thb!]
    \centering
    {\scriptsize (a)}
    \includegraphics[width=0.4\textwidth]{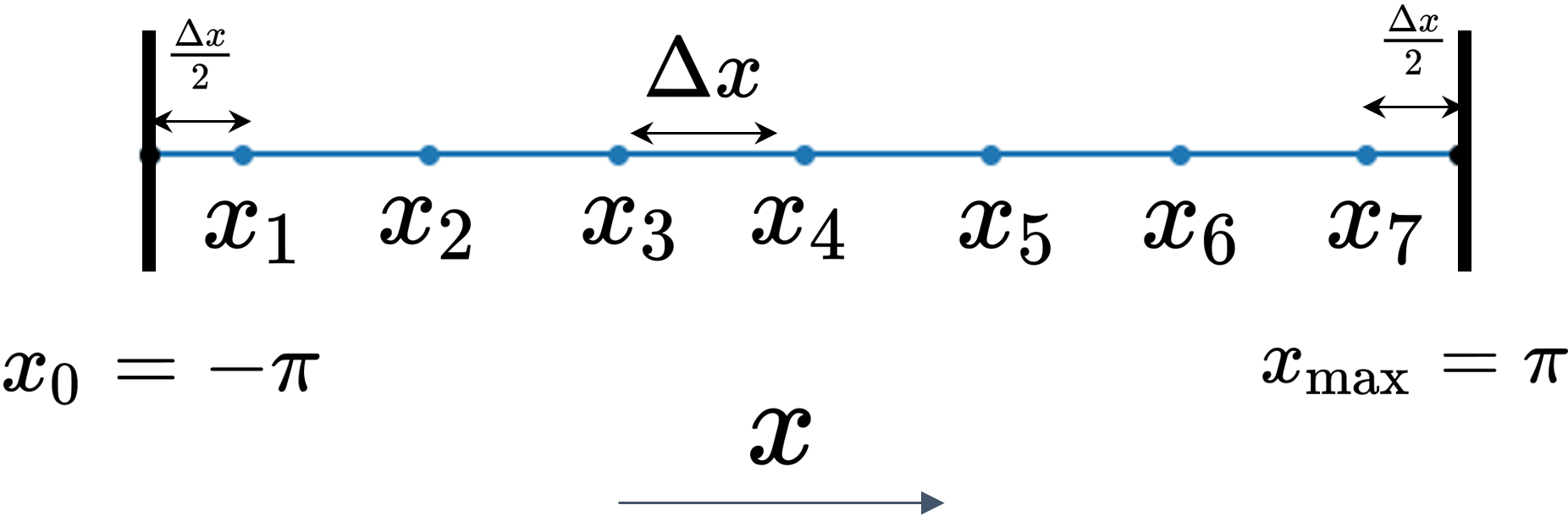}
    {\scriptsize (b)}     \includegraphics[width=0.2\textwidth]{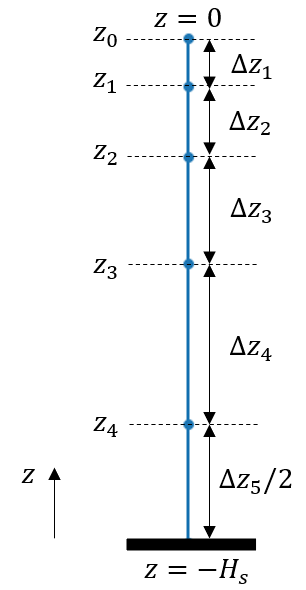}
    \caption{(a) Visual example of the cell-centered spatial grid in the $x-$direction for $N=7$. The nodes are spaced by $\Delta x$, except the the first and last grid point, which are spaced $\Delta x/2$ from the boundaries $x_0$ and $x_{\rm max}$, respectively. (b) Example of the nonuniform grid in the $z-$direction for $p=5$. Here, the spacing between grid points increases by a factor of $1.5$ at each increment.}
    \label{fig:thicksub_x_grid}
\end{figure}

We define the cell-centered spatial grid in the $x$-direction, used for both film and substrate:
\begin{align}
         x_i &= x_0 + \Delta x \left( i - 1/2 \right), \quad i=1,\ldots, N, \qquad
     \Delta x = \frac{\left(x_{\rm max} - x_0 \right)}{N}, \label{thicksub_x_grid}
\end{align}
where $N$ is the number of grid points in the $x$-direction, and the lateral boundaries are $x_0=-\pi$ and $x_{\rm max}=\pi$. An example of the spatial grid is given in Figure \ref{fig:thicksub_x_grid}(a), when $N=7$.

Similarly, let $p$ be the number of grid points in the $z$-direction (relevant only in the substrate). To reduce the computational expense, we use a nonuniform grid in the substrate, with grid points $\{z_k\}$ and variable step sizes $\{\Delta z_k\}$, $k=0,1,\ldots,p-1$, where the step sizes are taken to be geometric, with ratio $r$,
\begin{align}
    \Delta z_{k+1} = r \Delta z_k, \qquad k=1,\ldots, p-1.
\end{align}
Figure \ref{fig:thicksub_x_grid}(b) shows an example when $p=5$ and $r=1.5$ (the value of $r$ used in all results). The point $z_0=0$ is always fixed at the liquid-solid interface, $z=0$, and $z_{p-1}$ is the final grid point, which lies a distance $\Delta z_{p}/2$ above $z=-H_{\rm s}$. 
We then fix the first (minimum) step size, $\Delta z_{\rm min}=\Delta z_1$ to ensure that $\{\Delta z_k\}$, $k=1,2,\ldots, p-1$, gives the desired geometric partition of $[-H_{\rm s},0]$,
\begin{align}
    \Delta z_{\rm min} = H_{\rm s}  \left( \sum\limits_{k=1}^{p-1} r^{k-1} + \frac{1}{2} r^{p-1} \right)^{-1}.
\end{align}
With $\Delta z_{\rm min}$ defined, we can consistently define the sequence of step sizes and grid:
\begin{align}
    \Delta z_{k} &= \Delta z_{\rm min} r^{k-1}, \qquad &&k=1,\ldots,p, \\
    z_k &= z_{k-1}-\Delta z_{k}, \qquad &&k=1, \ldots, p-1. \label{thicksub_z_grid}
\end{align}
We next proceed with the solution methods for the underlying equations. For simplicity, we begin with the solution scheme for Eq.~\eqref{thicksub_asymptotic_sub_eqn}. We define
\begin{align}
     S_{i,k}^{n} \approx T_{\rm s}(x_i,z_k,t_n), \quad i=1,2,\ldots, N, \quad k=0,1,\ldots, p-1,
\end{align}
to be a discrete approximation of substrate temperature, $T_{\rm s}$, on the spatial grid given above. First, we apply a Crank-Nicolson time-stepping scheme, which takes the discrete form
\begin{align}
    \frac{S_{i,k}^{n+1} - S_{i,k}^{n}}{\Delta t} &= \frac{1}{2} f_{i,k}^{n+1} + \frac{1}{2}f_{i,k}^n, \quad i = 1, \ldots, N, \qquad k=1,\ldots, p-1, \label{thicksub_discrete_system1}
\end{align}
where $f_{i,k}(S_{i,k-1}, S_{i,k}, S_{i,k+1}) \approx {\rm Pe}_{\rm s}^{-1} \partial_z \left( \kappa(T_{\rm s}) \partial_z T_{\rm s} \right)\vert_{(x,z)=(x_i,z_k)}$ is a nonlinear function of $S_{i,k-1}, S_{i,k}$, and $S_{i,k+1}$. For the remainder of the section, we suppress the subscript $i$ on $S_{i,k}$ and $f_{i,k}$, for simplicity. For completeness, we note that $f_{k}$ can be approximated as follows,
\begin{align}
     \partial_z &\left( \kappa(T_{\rm s}) \partial_z T_{\rm s} \right) = \kappa(T_{\rm s}) \partial_z^2 T_{\rm s}+ \kappa'(T_{\rm s}) \left( \partial_z T_{\rm s} \right)^2, \label{thicksub_substrate_product_rule} \\
         \partial_z &\left( \kappa(T_{\rm s}) \partial_z T_{\rm s} \right)\vert_{z=z_k} \approx A_k S_{k-1} + B_k S_k + C_k S_{k+1} + D_k \left( S_{k-1} - S_{k+1} \right)^2, \label{thicksub_discrete_zsub_derivative} \\
    A_k &= \frac{2\kappa(S_k)}{\Delta z_k \left( \Delta z_{k} + \Delta z_{k+1} \right)}, \quad B_k = \frac{-2\kappa(S_k)}{\Delta z_k \Delta z_{k+1}}, \\ C_k &= \frac{2\kappa(S_k)}{\Delta z_{k+1} \left( \Delta z_{k} + \Delta z_{k+1} \right)}, \quad D_k = \frac{\kappa'(S_k)}{\left( \Delta z_k + \Delta z_{k+1} \right)^2},
\end{align}
where each equation is applied for a fixed $i$, $\kappa'(S_k) = d\kappa(S_k)/dS_k$, and $k=1,2,\ldots, p-1$. 

The cases $k=1$ and $k=p-1$ in Eq.~\eqref{thicksub_discrete_zsub_derivative} involve unknowns $S_0$ and $S_p$, which are determined by discretizing the boundary condition at $z=0$ (Eq.~\eqref{thicksub_cont_temp_nd}) and at $z=-H_{\rm s}$ (Eq.~\eqref{thicksub_Bi_boundary_condition}), respectively. Since $z_0 =0$, $S_0$ is simply set to the film temperature, $S_0 = T_{i}^{n} \approx T_{\rm f}(x_i,t_n)$ for each $i$. The boundary condition given by Eq.~\eqref{thicksub_Bi_boundary_condition} is discretized as
\begin{align}
    \kappa\left( \frac{S_{p-1}+S_p}{2} \right) \left( \frac{S_{p-1}-S_p}{\Delta z_p} \right) = {\rm Bi} \left( \frac{S_{p-1}+S_p}{2} - T_{\rm a} \right), \label{thicksub_discrete_Bi_eqn}
\end{align}
which is a nonlinear equation for the unknown $S_p$ to be solved  at each node $x_i$. To solve Eq.~\eqref{thicksub_discrete_Bi_eqn}, we use a Newton method, although any convergent iterative method would suffice. 

Next, we assume that the substrate temperature at time $t_{n+1}$ can be written as
\begin{align}
    S_k^{n+1} = S_k^* + w_k, \label{thicksub_sk_guess_eqn}
\end{align}
where $S_k^*$ is the guess to the solution at time $t_{n+1}$ and $w_k$ is a correction to that guess, which we call a Newton correction in what follows to avoid confusion. Then, $f$ is linearized around the guess:
\begin{align}
    f_k^{n+1} = f_k(S_k^{n+1}) = f_k(S_k^* + w_k) \approx f_k(S_k^*) +  w_j \frac{\partial f_k}{\partial S_j}\vert_{S_j=S_j^*}, \label{thicksub_fk_eqn}
\end{align}
where $k=1,\ldots,p-1$, and $\partial f_k/\partial S_j\vert_{S_j=S_j^*}$ are the components of the Jacobian, denoted $F_{k,j}=\partial f_k/\partial S_j\vert_{S_j=S_j^*}$, evaluated at the guess for the next temperature $S_j^*$. Equation~\eqref{thicksub_discrete_system1} is then linearized by plugging in Eqs.~\eqref{thicksub_sk_guess_eqn},~\eqref{thicksub_fk_eqn}, leading to a linear system of equations for the correction $w_k$, where $S_k^*$ and $S_k^n$ are both known ($S_k^*$ is to be iterated):
\begin{align}
    \sum\limits_{j=1}^{p-1} \left( \delta_{k,j} - \frac{1}{2} \Delta t F_{k,j} \right) w_j = R_k, \qquad k=1,\ldots, p-1, \label{thicksub_linearized_sub_eqn}
\end{align}
where $\delta_{k,j}$ is the Kronecker delta, and the right-hand side is
\begin{align}
    R_k = S_k^{n} - S_k^* + \frac{1}{2} \Delta t f_k(S_k^*) + \frac{1}{2} \Delta t f_k(S_k^{n}), \qquad k=1,\ldots, p-1. \label{thicksub_R_eqn}
\end{align}
For simplicity, we abbreviate Eq.~\eqref{thicksub_linearized_sub_eqn} as $\left(\mathbf{A}\mathbf{w}=\mathbf{R}\right)_i$ with the understanding that each $(p-1)\times (p-1)$ linear system is to be solved for each $x_i$. Solving Eq.~\eqref{thicksub_linearized_sub_eqn} completes one step of the iteration. Next, we check that $|w_k / S_k^*|<tol$ for all $k$. If yes, the iteration is finished, and $S_k^* + w_k$ becomes the substrate temperature at time $t_{n+1}$ for each $k=1,2,\ldots,p-1$, namely $S_k^{n+1}=S_k^* + w_k$.  If not, the iteration is completed until the specified convergence criterion is reached. We use $tol = 10^{-9}$.

Next we describe the solution mechanism for film temperature, Eq.~\eqref{thicksub_asymptotic model1}. First, we define the approximation for film temperature and thickness by
\begin{align}
    T_{i}^{n} \approx T_{\rm f}(x_i,t_n), \quad h_i^{n}\approx h(x_i,t_n), \quad i=1,\ldots,N.
\end{align}
Next, for compactness, we define the following expressions
\begin{align}
    \Psi_i^{n} &= \frac{1}{{\rm Pe}_{\rm f}} \left[ \delta_x^2 T_i^{n} + \left( \frac{\delta_x h_i^n}{h_i^n}\right) \delta_x T_{i}^{n} \right], \label{thicksub_2d_psi_eqn} \\
    G_i^{n} &= -\frac{\mathcal{K}}{ {\rm Pe}_{\rm f} h_i^n} \left[\kappa(S_0^n)\delta_z^+( S_0^n ) \right]_i ,  \label{thicksub_2d_G_eqn}  
\end{align}
where $\left[\kappa(S_0^n)\delta_z^+( S_0^n )\right]_i \approx \kappa(T_{\rm s})\partial T_{\rm s}/\partial z \vert_{(x,z,t)=(x_i,0,t_n)}$ is an approximation of the heat flux along the liquid-solid interface, $z=0$, at node $x_i$ and time $t_n$, which we define as
\begin{align}
    &\kappa(S_0^n)\delta_z^+( S_0^n ) = \kappa(S_{0}^{n})\left(a_0 S_0^{n} + b_0 S_1^{n} + c_0 S_2^{n} \right), \\
    a_0 &= \frac{2\Delta z_1 + \Delta z_2}{\Delta z_1\left( \Delta z_1 + \Delta z_2 \right)}, \quad b_0 = - \left( \frac{1}{\Delta z_1} + \frac{1}{\Delta z_2} \right), \quad c_0 = \frac{\Delta z_1}{\Delta z_2 \left( \Delta z_1 + \Delta z_2 \right)}.
\end{align}
The second-order central difference approximations of $\partial_x, \partial_x^2$ are defined as $\delta_x$ and $\delta_x^2$, respectively, and are given by
\begin{align}
    \delta_x T_i=\frac{T_{i+1}-T_{i-1}}{2\Delta x}, \qquad
\delta_x^2 T_{i} = \frac{T_{i+1}-2T_{i}+T_{i-1}}{\Delta x^2}, \qquad i=1,\ldots,N, \nonumber
\end{align}
where $T_0, T_{N+1}$ can be written in terms of $T_1$ and $T_{N}$, respectively, by solving discretized versions of
Eq.~\eqref{thicksub_insulating_cond} at the lateral boundaries, $x=\pm \pi$, 
\begin{align}
    \partial_x \left( T_{\rm f} \right)\vert_{x=-\pi} \approx \frac{T_1-T_0}{\Delta x} = 0, \qquad \partial_x \left( T_{\rm f} \right)\vert_{x=\pi} \approx \frac{T_{N+1}-T_{N}}{\Delta x} = 0. \label{thicksub_lateral_bc_eqn}
\end{align}
\begin{figure}[H]
    \centering
    \includegraphics[width=0.45\textwidth]{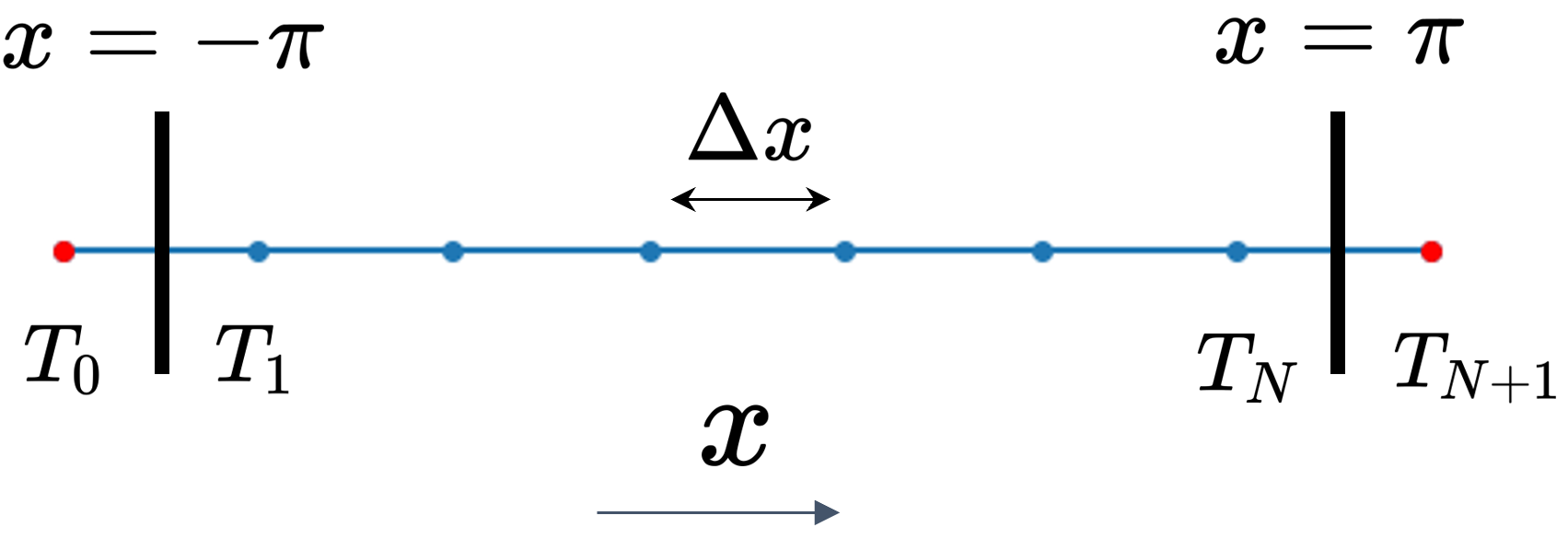}
    \caption{Depiction of the discretized film temperature adjacent to the lateral boundaries, $x=\pm \pi$ (black, vertical bars). The {\color{blue} blue} nodes represent the grid $x_1,x_2,\ldots,x_N$ with spacing $\Delta x$. The film temperature at the first and last interior grid points, near $x=-\pi, \pi$, are $T_1,T_N$, and $T_0,T_{N+1}$ represent ghost points, located at $x=\pm \left( \pi + \Delta x/2 \right)$ ({\color{red} red} nodes).}
    \label{fig:thicksub_lateralboundaries_grid_schematic}
\end{figure}
Figure \ref{fig:thicksub_lateralboundaries_grid_schematic} shows the spatial grid in the $x$-direction. 
The {\color{red} red} nodes represent ghost points with temperatures $T_0,T_{N+1}$. 
By solving Eq.~\eqref{thicksub_lateral_bc_eqn}, we obtain $T_1=T_0$ and $T_{N+1}=T_N$.

Now, to solve Eqs.~\eqref{thicksub_asymptotic model1} and \eqref{thicksub_asymptotic_sub_eqn} for $T_{\rm f}$ and $T_{\rm s}$, we use a predictor-corrector Runge-Kutta/Crank-Nicolson scheme combined with a Newton method as described above. In what follows, hatted quantities denote those found in the predictor phase, whereas those without hats are determined in the corrector phase. In the predictor phase, one finds intermediate ``predicted" film and substrate temperatures $(\hat{T}_i^{n+1},\hat{S}_{k}^{n+1})$. In the corrector phase, one uses the intermediate variables to find corrected film and substrate temperatures $(T_i^{n+1},S_{k}^{n+1})$. In both cases, the substrate temperature is found by solving the linear systems given by Eq.~\eqref{thicksub_linearized_sub_eqn} for $\hat{w}$ or $w$. In the former case, the nonlinear system that is linearized is Eq.~\eqref{thicksub_discrete_system1} with $\hat{S}_{k}^{n+1}$ in place of $S_{k}^{n+1}$ and with $f_{k}^{n+1}$ replaced by $\hat{f}_k^{n+1}(\hat{S}_{k-1}^{n+1},\hat{S}_k^{n+1},\hat{S}_{k+1}^{n+1})$. In the predictor phase, we use a forward-Euler scheme to deal with $G_i^n$: 
\begin{align}
\frac{\hat{T}_{i}^{n+1}-T_{i}^n}{\Delta t}&=\frac{1}{2} \left[ \hat{\Psi}_i^{n+1} + \Psi_i^{n} \right]  +  G_i^{n} + \overline{Q}_{i}^{n+1/2}, \quad i= 1, \ldots, N,  \label{thicksub_film_temp_prediction} \\
\left(\hat{\mathbf{A}}\hat{\mathbf{w}}=\hat{\mathbf{R}}\right)_i, &\quad i = 1, \ldots, N, \label{thicksub_discrete_system2}
\end{align}
where $\overline{Q}_i^{n+1/2}=(\overline{Q}_i^n+\overline{Q}_i^{n+1})/2$, and $\hat{\Psi}_i^{n+1}$ is found by substituting $\hat{T}_i^{n+1}$ in place of $T_{i}^{n+1}$ in Eq.~\eqref{thicksub_2d_psi_eqn}. Similarly, the components of $\hat{\mathbf{w}}$ are related to the predicted substrate temperature, $\hat{S}_k^{n+1}$, via Eq.~\eqref{thicksub_sk_guess_eqn} with appropriate substitution.

Solving Eq.~\eqref{thicksub_film_temp_prediction} provides the predicted temperature $\hat{T}_i^{n+1}$. The linearized system given by Eq.~\eqref{thicksub_discrete_system2} (where $\hat{\mathbf{A}},\hat{\mathbf{R}}$ are found using $\hat{S}_k^*$, the guess to $\hat{S}_k^{n+1}$ and $\hat{S}_k^{n}$ in Eqs. \eqref{thicksub_linearized_sub_eqn} and \eqref{thicksub_R_eqn}) is solved iteratively for $\hat{\mathbf{w}}$ and each $i$. The predictor phase amounts to solving one linear system of size $N$ for the film and $N$ linear systems of size $p-1$ for the substrate. More importantly, the solution to Eq.~\eqref{thicksub_discrete_system2} in the predictor phase gives us an approximation of substrate temperature, $\hat{S}_{i,k}^n$, so that we can calculate a prediction to the heat flux at the liquid-solid interface, $\hat{G}_i^n$. We then correct the temperature predictions by using a second-order Runge-Kutta method on $G_i^n$ using $\hat{G}_i^n$:
\begin{align}
\frac{T_{i}^{n+1}-T_{i}^n}{\Delta t}&=\frac{1}{2} \left[ \Psi_i^{n+1} + \Psi_i^{n} \right]  + \frac{1}{2} \left( G_i^{n} + \hat{G}_i^{n} \right) + \overline{Q}_{i}^{n+1/2}, \quad i = 1, \ldots, N, \label{thicksub_film_temp_correction} \\
\left(\mathbf{A}\mathbf{w}=\mathbf{R}\right)_i, &\quad i = 1, \ldots, N. \label{thicksub_sub_temp_correction}
\end{align}

Next we describe the numerical scheme for film thickness, $h(x,t)$. First, we use the Crank-Nicolson scheme to discretize Eq.~\eqref{thicksub_thin_film} in time. The resulting nonlinear system of equations is given by Eq.~\eqref{thicksub_discrete_thin_film}, where $D_i(t)=D(h(x_i,t_n))\approx D_i^{n}$ is a second-order accurate spatial discretization of the  derivative of flux,
\begin{align}
    D = -\partial_x \cdot \left[\frac{1}{\mathcal{M}} \left(  h^3  \partial_x \left( \Gamma \partial_{xx} h + \Pi(h) \right)  \right) \right].
\end{align}
Following the procedure implemented for solving Eq.~\eqref{thicksub_discrete_system1} we apply a Newton method, first linearizing the film thickness around a guess, $h_i^*$, and solving a resultant linear system for the Newton correction to the guess,
\begin{align}
    h_i^{n+1} &= h_i^* + q_i^*, \quad i=1,\ldots, N, \\
    \mathbf{A}_h \mathbf{q} &= \mathbf{R}_h, \label{thicksub_thin_film_linear_system}
\end{align}
where $(\mathbf{A}_h)_{i,j} = \partial D_i/\partial h_j\vert_{h_j = h_j^*}$ are the components of the Jacobian, $\mathbf{q}$ is the Newton correction vector for $\mathbf{h}$, and $\mathbf{R}_h$ is the remainder, whose components, $(R_h)_i$, $i=1,2,\ldots, N$, are analogous to Eq.~\eqref{thicksub_R_eqn}:
\begin{align}
    \left( R_h\right)_i = h_i^n - h_i^* + \frac{1}{2}\Delta t D_i^* + \frac{1}{2}\Delta t D_i^n,
\end{align}
and where $D_i^* \approx D(h_i^*)$ is an approximation of the flux with the guess.
For more details regarding the 2D solution mechanism for $h$, we refer the reader to Kondic~\cite{siam}.

\begin{figure}[H]
    \centering
    \includegraphics[width=\textwidth]{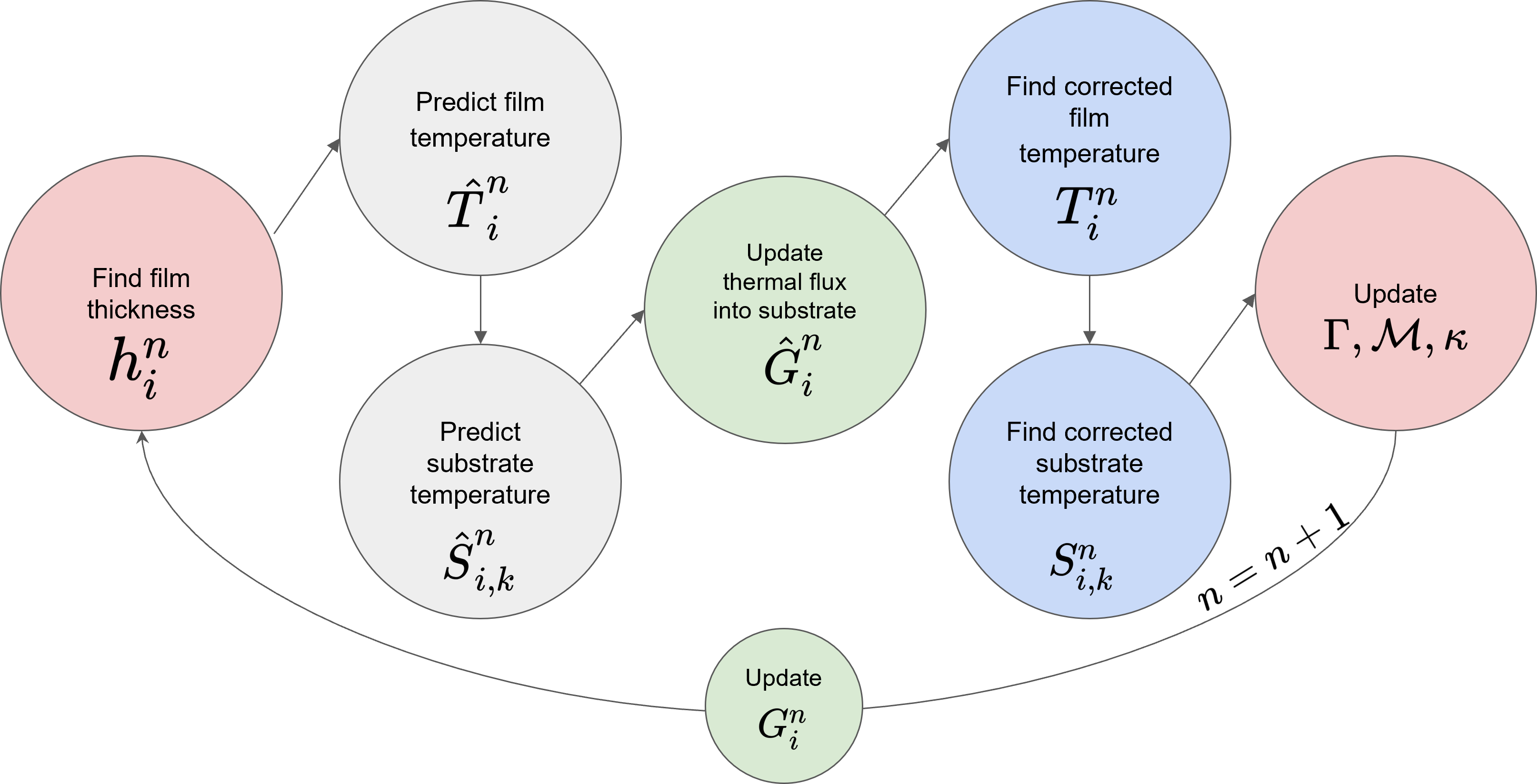}
    \caption{Flowchart for the 2D numerical method used to solve for $h(x,t)$, $T_{\rm f}(x,t)$ and $T_{\rm s}(x,z,t)$.}
    \label{thicksub_fig:2d_flowchart}
\end{figure}

Figure~\ref{thicksub_fig:2d_flowchart} shows a flowchart of the solution process for finding film thickness, film temperature, and substrate temperature. {\color{red} Red} circles indicate the beginning and end of a time-step iteration. {\color{gray} Gray} circles indicate the prediction step for heat conduction and the {\color{blue} blue} circles represent the correction step. The {\color{darkseagreen} green} circles represent intermediate stages where the thermal flux into the substrate is updated. First, $h$ is found at time $t_n$ by solving Eq.~\eqref{thicksub_thin_film_linear_system} for every spatial node $x_i$. That value of $h$ is then used to solve Eq.~\eqref{thicksub_film_temp_prediction} for a prediction of the film temperature, $\hat{T}_{i}^{n}$. That film temperature is then used to solve for the predicted substrate temperature, $\hat{S}_{i,k}^{n}$ via Eq.~\eqref{thicksub_discrete_system2}. The thermal flux at the liquid-solid interface, $\hat{G}_i^n$, is then updated using $\hat{S}_{i,k}^{n}$ in Eq.~\eqref{thicksub_2d_G_eqn}. These temperature predictions are then corrected using Eqs.~\eqref{thicksub_film_temp_correction} and \eqref{thicksub_sub_temp_correction}. Surface tension, $\Gamma$, viscosity, $\mathcal{M}$, and substrate thermal conductivity at $z=0$, $\kappa(S_{i,0}^{n})$, are then updated. Finally, until the desired end time is reached, time is incremented, and $G_i^{n}$ is updated using $S_{i,k}^n$ in Eq.~\eqref{thicksub_2d_G_eqn}.

\subsection{3D Numerical schemes}\label{thicksub_sect:3d_numerics}

Here, we consider the numerical scheme to solve the full 3D versions of Eqs.~\eqref{thicksub_thin_film},~\eqref{thicksub_asymptotic model1}--\eqref{thicksub_asymptotic_sub_eqn}, where $h=h(x,y,t),T_{\rm f}=T_{\rm f}(x,y,t)$, and $T_{\rm s}=T_{\rm s}(x,y,z,t)$.
Since $y$-dependence is now included (see Fig.~\ref{thicksub_fig:schematic}),
the complexity of the numerical problems are, as a minimum, increased by a factor of $N$ for each set of equations. This creates a computational challenge, which makes serial CPU computing prohibitively slow. Parallel computing is a much more practical route. 
%\note[LK]{sorry, I don't understand the following sentence. It may need to be split into 2-3 shorter ones.}
For example, the finite-difference method discretization of PDEs often leads to tri-diagonal linear systems (such as Eq.~\eqref{thicksub_discrete_system2}). In these cases, either the formation of the matrix/vector system, or
the solution method itself, can be parallelized. Parallelization of the matrix/vector system may be done, for example, by defining the value of each element in parallel. Solving the tri-diagonal linear systems in parallel is less trivial since the Thomas Algorithm, typically used for such problems, is naturally sequential. To compensate, parallel cyclic reduction methods have been proposed that trade complexity for speed and prove superior to the traditional Thomas algorithm for many problems \cite{Zhang2010_GPU}. We use a simpler approach, however, by solving each linear system in parallel rather than parallelization of the solver (details following below). 

Parallel computing with multi-node systems and multi-core processors is also used in scientific computing but is resource-limited by the number of cores available per CPU. GPUs, on the other hand, have thousands of ``cores" available for computing and allow the programmer many more degrees of freedom in parallelization~\cite{Sanders}. Various CUDA algorithms have been developed for solving penta-diagonal systems \cite{Lam2019,Gloster2019}, for example, which often arise from 4th order PDEs such as Eq.~\eqref{thicksub_thin_film}. Recent
work~\cite{Lam2019} described a GPU-based code that can be used to solve thin film problems, finding a near $150$ times speed up over similar CPU-based code for certain domain sizes. The present work uses an extension of that code, which also incorporates thermal effects with CUDA, described below.

The remainder of the section is structured as follows. First, we define the 3D spatial grid. Then, we describe the solution methodology for computing temperatures, both in the film and in the substrate.  Finally, we conclude with the solution mechanism for film thickness.
We focus mainly here on the aspects of the implementation that are specific to the 3D geometry.  

The $x$-component of the spatial grid is given by Eq.~\eqref{thicksub_x_grid} and the $z$-component of the substrate grid by Eq.~\eqref{thicksub_z_grid}. 
We similarly introduce the $y$-component of the spatial grid,
\begin{align}
         y_j &= y_0 + \Delta y \left( j - 1/2 \right), \quad j=1,\ldots, M, \qquad
     \Delta y = \frac{\left(y_{\rm max} - y_0 \right)}{M}, \label{thicksub_y_grid}
\end{align}
where $M$ is the number of grid points in the $y$-direction. Therefore, the film grid consists of $N\times M$ interior nodes $\{(x_i,y_j),\, i=1,2,\ldots,N,\, j=1,2,\ldots,M\}$. In the substrate there are $N\times M \times p$ nodes $(x_i,y_j,z_k)$.

Similarly to Appendix~\ref{thicksub_sect:2d_numerics}, we define 
\begin{align}
    T_{i,j}^{n}\approx T_{\rm f}(x_i,y_j,t_n), \quad S_{i,j,k}^{n} \approx T_{\rm s}(x_i,y_j,z_k,t_n), \quad h_{i,j}^{n} \approx h(x_i,y_j,t_n),
\end{align}
as approximations to the film and substrate temperatures, and film thickness. The predictor/corrector solution methodology from Appendix~\ref{thicksub_sect:2d_numerics} is applied once more, except Eq.~\eqref{thicksub_asymptotic model1} now requires an alternating-direction implicit (ADI) method to achieve second-order accuracy. Similarly to Appendix~\ref{thicksub_sect:2d_numerics}, we begin with a predictor step to find $\hat{\mathbf{w}}$ and $(\hat{T}_{i,j}^{n+1},\hat{S}_{i,j,k}^{n+1})$:
\begin{align}
\frac{T_{i,j}^* -T_{i,j}^n}{\Delta t} &= \frac{1}{2} X_{i,j}^* + \frac{1}{2}Y_{i,j}^n + \frac{1}{2}G_{i,j}^n + \frac{1}{2}\overline{Q}_{i,j}^{n+1/2},\label{thicksub_3d_adi1}  \\
\frac{\hat{T}_{i,j}^{n+1} - T_{i,j}^*}{\Delta t} &= \frac{1}{2}X_{i,j}^* + \frac{1}{2}\hat{Y}_{i,j}^{n+1} +  \frac{1}{2}G_{i,j}^n + \frac{1}{2}\overline{Q}_{i,j}^{n+1/2}, \label{thicksub_3d_adi2} \\
\left(\hat{\mathbf{A}} \hat{\mathbf{w}}=\hat{\mathbf{R}}\right)_{i,j},&  \label{thicksub_discrete_system3}
\end{align}
where
\begin{align}
X_{i,j} &= {\rm Pe}_{\rm f}^{-1} \left[ \delta_x^2 T_{i,j} + \left( \frac{\delta_x h_{i,j}}{h_{i,j}} \right) \delta_x T_{i,j} \right] , \label{thicksub_X_eqn} \\
Y_{i,j} &= {\rm Pe}_{\rm f}^{-1} \left[ \delta_y^2 T_{i,j} + \left( \frac{\delta_y h_{i,j}}{h_{i,j}} \right) \delta_y T_{i,j} \right], \label{thicksub_Y_eqn} \\
\overline{Q}_{i,j}^{n+1/2} &= \frac{\overline{Q}_{i,j}^n + \overline{Q}_{i,j}^{n+1}}{2}, \\
G_{i,j}^{n} &= -\frac{\mathcal{K}}{ {\rm Pe}_{\rm f} h_{i,j}^n} \left[\kappa(S_0^n)\delta_z^+( S_0^n )\right]_{i,j},
\end{align}
$h_{i,j}^{n} \approx h(x_i,y_j,t_n)$, $i=1, \ldots, N$, $j=1,\ldots,M$, and $\kappa(S_0^n)\delta_z^+( S_0^n )$ approximates the heat flux at the interface $z=0$ and is given in Appendix~\ref{thicksub_sect:2d_numerics}. The term $T_{i,j}^*$ is the solution at an intermediate step between times $t_n, t_{n+1}$, and $\hat{\mathbf{A}},\hat{\mathbf{R}}$ are defined as in Appendix \ref{thicksub_sect:2d_numerics}, but with the extra index $j$. The solution for $h$ is only found at times $t_n$ and $t_{n+1}$, so we approximate $h$ at the intermediate step as
\begin{align}
    h_{i,j}^* = \frac{h_{i,j}^{n} + h_{i,j}^{n+1}}{2}.
\end{align}
Equation~\eqref{thicksub_3d_adi1} yields $M$ linear systems of equations of size $N$. Similarly, Eq.~\eqref{thicksub_3d_adi2} yields $N$ linear systems of equations of size $M$. Since the ADI method treats one variable explicitly and the other implicitly, both Eqs.~\eqref{thicksub_3d_adi1} and \eqref{thicksub_3d_adi2} are solved in parallel for each $j$ and each $i$, respectively (the formation of the linear system is also parallelized; for example, $T_{i,j}^*-(\Delta t/2) X_{i,j}^*$ for $j$ fixed and $i=1,\ldots,N$ are the components of the $N\times N$ matrix in Eq.~\eqref{thicksub_3d_adi1}, which are all found simultaneously). The 3D numerical code used here is freely available \cite{GADIT_Thermal}. 

Equation~\eqref{thicksub_discrete_system3} is the 3D analog of Eq.~\eqref{thicksub_discrete_system2}, but now there are $N\times M$ linear systems of equations of size $p-1$. Since Eq.~\eqref{thicksub_asymptotic_sub_eqn} only involves $z$-derivatives, Eq.~\eqref{thicksub_discrete_system3} is trivially parallelized for each $i$ and $j$. Since the solution of Eq.~\eqref{thicksub_discrete_system3} is iterative, careful consideration of the size of domains and the relation to memory performance is crucial. In our computations, $p$ is relatively small in comparison to $N$ and $M$ so that for each $i$ and $j$ both the matrix and vector of the linear system (of size $p-1$) can fit on shared memory on the device (GPU), which is known to be computationally advantageous over the use of global memory \cite{Sanders}.  

Next, we correct the predictor step using the Runge-Kutta method on $G_{i,j}$,
\begin{align}
\frac{T_{i,j}^* -T_{i,j}^n}{\Delta t} &= \frac{1}{2}X_{i,j}^* + \frac{1}{2}Y_{i,j}^n + \frac{1}{4} \left( G_{i,j}^n + \hat{G}_{i,j}^n \right) + \frac{1}{2}\overline{Q}_{i,j}^{n+1/2}, \label{thicksub_3d_adi1_2}  \\
\frac{T_{i,j}^{n+1} - T_{i,j}^*}{\Delta t} &= \frac{1}{2}X_{i,j}^* + \frac{1}{2}Y_{i,j}^{n+1} + \frac{1}{4} \left( G_{i,j}^n + \hat{G}_{i,j}^n \right) + \frac{1}{2}\overline{Q}_{i,j}^{n+1/2}, \label{thicksub_3d_adi2_2} \\
\left(\mathbf{A}\mathbf{w}=\mathbf{R}\right)_{i,j},
\end{align}
where $i=1, \ldots, N$, and $j=1,\ldots,M$. We note that although the repetitive nature of the predictor-corrector scheme may appear as a performance bottleneck, in our implementation the results from the predictor phase are stored to global memory and imported into 
the corrector step to speed up the computations. 

Next, we briefly describe the solution mechanism for film thickness $h$. Now, $h=h(x_i,y_j,t_n)$ but the approach is very similar to that of Appendix~\ref{thicksub_sect:2d_numerics}. First we define the divergence of the flux
\begin{align}
    D = -\bnabla_2 \cdot \left[\frac{1}{\mathcal{M}} \left(  h^3  \bnabla_2 \left( \Gamma \bnabla_2^2 h + \Pi(h) \right)  \right) \right],
\end{align}
and define $D_{i,j}^{n}$ to be a second-order spatial discretization of $D$. Equation~\eqref{thicksub_thin_film} can then be written as
\begin{align}
    \frac{h_{i,j}^{n+1}-h_{i,j}^n}{\Delta t} = \frac{1}{2}D_{i,j}^{n+1} + \frac{1}{2}D_{i,j}^{n} , \quad i=1,\ldots, N,\; j=1,\ldots, M. \label{thicksub_thin_film_discrete_3d}
\end{align}
Equation~\eqref{thicksub_thin_film_discrete_3d} is linearized and a Newton's method is used to iterate guesses to the film thickness at time $t_{n+1}$. In contrast to the 2D case, $D$ now involves derivatives with respect to $y$ as well as $x$. Therefore, the Newton's method is split into two separate linear systems of equations (one where $y$-derivatives are treated implicitly in time and one similarly for $x$-derivatives), and solved iteratively. The equations in general take the form
\begin{align}
    \mathbf{A}_{y,(l)} \mathbf{w}_h &= \mathbf{b}_{y,(l)}, \label{thicksub_3d_gpu1} \\
    \mathbf{A}_{x,(l)} \mathbf{v} &= \mathbf{w}_h, \label{thicksub_3d_gpu2} \\
    \mathbf{h}_{(l+1)}^{n+1} &= \mathbf{h}_{(l)}^{n+1} + \mathbf{v},
\end{align}
where $(l)$ represents iteration number, $\mathbf{h}$ represents the array of values $h_{i,j}$, $\mathbf{w}_h$ is an intermediate step, $\mathbf{v}$ is an array of corrections to the guess $\mathbf{h}_{(l)}^{n+1}$, $\mathbf{A}_{y,(l)}, \mathbf{A}_{x,(l)}$ are matrices whose components are found using pure $y$- and $x$-derivative terms, respectively, and $\mathbf{b}_{y,(l)}$ is a vector (containing flux discretizations), which we omit for brevity. For details regarding these terms we refer the reader to the work of Lam \textit{et al.} \cite{Lam2019}. Notably, Eqs.~\eqref{thicksub_3d_gpu1} and \eqref{thicksub_3d_gpu2} are penta-diagonal systems, which can be solved in parallel. In the former, $N$ linear systems of equations of size $M\times M$ are solved simultaneously, while in the latter, 
the same is done for $M$ linear systems of size $N\times N$.

The film thickness is again coupled to film temperature through the material parameters, film temperature is coupled to thickness via Eqs.~\eqref{thicksub_X_eqn} and ~\eqref{thicksub_Y_eqn}, and substrate temperature to film temperature via the interface $z=0$. The solution order is identical to that of Appendix~\ref{thicksub_sect:2d_numerics}, solving first for $h$ and then $T_{\rm f}$ and $T_{\rm s}$ using a predictor-corrector method.

\bibliography{films}

\begin{thebibliography}{10}

\bibitem{oron_rmp97}
A.~Oron, S.~H. Davis, and S.~G. Bankoff.
\newblock {Long-scale evolution of thin liquid films}.
\newblock {\em Rev. Mod. Phys.}, 69:931--980, 1997.

\bibitem{craster_rmp09}
R.~V. Craster and O.~K. Matar.
\newblock {Dynamics and stability of thin liquid films}.
\newblock {\em Rev. Mod. Phys.}, 81:1131, 2009.

\bibitem{Tseluiko_SIAM2007}
D.~Tseluiko and D.~T. Papageorgiou.
\newblock Nonlinear dynamics of electrified thin liquid films.
\newblock {\em SIAM J. Appl. Math}, 67:1310--1329, 2007.

\bibitem{Mema2020}
E.~Mema, L.~Kondic, and L.~J. Cummings.
\newblock {Dielectrowetting of a thin nematic liquid crystal layer}.
\newblock {\em Phys. Rev. E}, 103:032702, 2020.

\bibitem{chappell_o'dea_2020}
D.~J. Chappell and R.~D. O'Dea.
\newblock Numerical-asymptotic models for the manipulation of viscous films via
  dielectrophoresis.
\newblock {\em {J. Fluid Mech.}}, 901:A35, 2020.

\bibitem{Frolovskaya2008}
O.~A. Frolovskaya, A.~A. Nepomnyashchy, A.~Oron, and A.~A. Golovin.
\newblock {Stability of a two-layer binary-fluid system with a diffuse
  interface}.
\newblock {\em {Phys. Fluids}}, 20:1--17, 2008.

\bibitem{Naraigh2010}
L.~{\'{O}} N{\'{a}}raigh and J.~L. Thiffeault.
\newblock {Nonlinear dynamics of phase separation in thin films}.
\newblock {\em Nonlinearity}, 23:1559--1583, 2010.

\bibitem{Diez2021}
J.~A. Diez, A.~G. González, D.~A. Garfinkel, P.~D. Rack, J.~T. McKeown, and
  L.~Kondic.
\newblock Simultaneous decomposition and dewetting of nanoscale alloys: A
  comparison of experiment and theory.
\newblock {\em Langmuir}, 37:2575--2585, 2021.

\bibitem{Allaire_JPC2021}
R.~H. Allaire, L.~Kondic, L.~J. Cummings, P.~D. Rack, and M.~Fuentes-Cabrera.
\newblock {The Role of Phase Separation on Rayleigh-Plateau Type Instabilities
  in Alloys}.
\newblock {\em J. Phys. Chem. C}, 2021.

\bibitem{Thiele2013}
U.~Thiele, D.~V. Todorova, and H.~Lopez.
\newblock {Gradient dynamics description for films of mixtures and suspensions:
  Dewetting triggered by coupled film height and concentration fluctuations}.
\newblock {\em {Phys. Rev. Lett.}}, 111:1--5, 2013.

\bibitem{Davis2000}
S.~H. Davis and L.~M. Hocking.
\newblock {Spreading and imbibition of viscous liquid on a porous base. II}.
\newblock {\em Phys. Fluids}, 12:1646--1655, 2000.

\bibitem{Zadrazil2006}
A.~Zadra{\v{z}}il, F.~Stepanek, and O.~K. Matar.
\newblock Droplet spreading, imbibition and solidification on porous media.
\newblock {\em {J. Fluid Mech.}}, 562:1–33, 2006.

\bibitem{trice_prb07}
J.~Trice, D.~Thomas, C.~Favazza, R.~Sureshkumar, and R.~Kalyanaraman.
\newblock {Pulsed-laser-induced dewetting in nanoscopic metal films: Theory and
  experiments}.
\newblock {\em Phys. Rev. B}, 75:235439, 2007.

\bibitem{atena09}
A.~Atena and M.~Khenner.
\newblock Thermocapillary effects in driven dewetting and self assembly of
  pulsed-laser-irradiated metallic films.
\newblock {\em {Phys. Rev. B}}, 80:075402, 2009.

\bibitem{Saeki2011}
F.~Saeki, S.~Fukui, and H.~Matsuoka.
\newblock {Optical interference effect on pattern formation in thin liquid
  films on solid substrates induced by irradiative heating}.
\newblock {\em {Phys. Fluids}}, 23:112102, 2011.

\bibitem{Saeki2013}
F.~Saeki, S.~Fukui, and H.~Matsuoka.
\newblock {Thermocapillary instability of irradiated transparent liquid films
  on absorbing solid substrates}.
\newblock {\em {Phys. Fluids}}, 25:062107, 2013.

\bibitem{font2017}
F.~Font, S.~Afkhami, and L.~Kondic.
\newblock Substrate melting during laser heating of nanoscale metal films.
\newblock {\em Int. J. Heat Mass Transfer}, 113:237, 2017.

\bibitem{Seric_pof2018}
I.~Seric, S.~Afkhami, and L.~Kondic.
\newblock Influence of thermal effects on stability of nanoscale films and
  filaments on thermally conductive substrates.
\newblock {\em {Phys. Fluids}}, 30:012109, 2018.

\bibitem{rack_nano08}
Y.~F. Guan, R.~P. Pearce, A.~V. Melecho, D.~K. Hensley, M.~L. Simpson, and
  P.~D. Rack.
\newblock {Pulsed laser dewetting of nickel catalyst for carbon nanofiber
  growth}.
\newblock {\em Nanotechnology}, 19:235604, 2008.

\bibitem{Zhang2010}
S.~Zhang.
\newblock {\em {Nanostructured thin films and coatings : functional
  properties}}.
\newblock CRC Press, 2010.

\bibitem{atwater_natmat10}
H.A. Atwater and A.~Polman.
\newblock {Plasmonics for improved photovoltaic devices}.
\newblock {\em Nat. Mat.}, 9:205--213, 2010.

\bibitem{Makarov2016}
S.~V. Makarov, V.~A. Milichko, I.~S. Mukhin, I.~I. Shishkin, D.~A. Zuev, A.~M.
  Mozharov, A.~E. Krasnok, and P.~A. Belov.
\newblock {Controllable femtosecond laser-induced dewetting for plasmonic
  applications}.
\newblock {\em Laser Photonics Rev.}, 10:91--99, 2016.

\bibitem{Hughes2017}
R.~A. Hughes, E.~Menumerov, and S.~Neretina.
\newblock {When lithography meets self-assembly: a review of recent advances in
  the directed assembly of complex metal nanostructures on planar and textured
  surfaces}.
\newblock {\em Nanotechnology}, 28:282002, 2017.

\bibitem{kondic_arfm_2020}
L.~Kondic, A.~G. Gonzalez, J.~A. Diez, J.~D. Fowlkes, and P.~Rack.
\newblock Liquid-state dewetting of pulsed-laser-heated nanoscale metal films
  and other geometries.
\newblock {\em Annu. Rev. Fluid Mech.}, 52:235--262, 2020.

\bibitem{allaire_jfm2021}
R.~H. Allaire, L.~J. Cummings, and L.~Kondic.
\newblock On efficient asymptotic modelling of thin films on thermally
  conductive substrates.
\newblock {\em {J. Fluid Mech.}}, 915:A133, 2021.

\bibitem{Dong_prf16}
N.~Dong and L.~Kondic.
\newblock Instability of nanometric fluid films on a thermally conductive
  substrate.
\newblock {\em Phys. Rev. Fluids}, 1:063901, 2016.

\bibitem{shklyaev12}
S.~Shklyaev, A.~A. Alabuzhev, and M.~Khenner.
\newblock Long-wave {M}arangoni convection in a thin film heated from below.
\newblock {\em Phys. Rev. E}, 85:016328, 2012.

\bibitem{batson2019}
W.~Batson, L.~J. Cummings, D.~Shirokoff, and L.~Kondic.
\newblock Oscillatory thermocapillary instability of a film heated by a thick
  substrate.
\newblock {\em {J. Fluid Mech.}}, 872:928–962, 2019.

\bibitem{Beerman2007}
M.~Beerman and L.~N. Brush.
\newblock {Oscillatory instability and rupture in a thin melt film on its
  crystal subject to freezing and melting}.
\newblock {\em {J. Fluid Mech.}}, 586:423--448, 2007.

\bibitem{oron00}
A.~Oron.
\newblock {Three dimensional nonlinear dynamics of thin liquid films}.
\newblock {\em Phys. Rev. Lett.}, 85:2108, 2000.

\bibitem{lang13}
A.~G. Gonzalez, J.~D. Diez, Y.~Wu, J.D. Fowlkes, P.~D. Rack, and L.~Kondic.
\newblock {Instability of Liquid Cu Films on a SiO$_2$ Substrate}.
\newblock {\em Langmuir}, 13:9378--9387, 2013.

\bibitem{metals_ref_book_2004}
W.F. Gale and T.C. Totemeier.
\newblock {\em Smithells Metals Reference Book (Eighth Edition)}.
\newblock Butterworth-Heinemann, 2004.

\bibitem{kaptay}
G.~Kaptay.
\newblock A unified equation for the viscosity of pure liquid metals.
\newblock {\em Zeitschrift für Metallkunde}, 96:24--31, 2005.

\bibitem{Lam2019}
M.~A. Lam, L.~J. Cummings, and L.~Kondic.
\newblock {Computing dynamics of thin films via large scale GPU-based
  simulations}.
\newblock {\em J. Comput. Phys.: X}, 2:100001, 2019.

\bibitem{cuda}
NVIDIA, Péter Vingelmann, and Frank~H.P. Fitzek.
\newblock Cuda, release: 10.2.89, 2020.

\bibitem{Powell_1966}
R.~W. Powell, C.~Y. Ho, P.~E. Liley, States United, Standards National
  Bureau~of, States United, and Commerce Department~of.
\newblock {\em Thermal conductivity of selected materials}.
\newblock U.S. Dept. of Commerce, National Bureau of Standards; for sale by the
  Superintendent of Documents, U.S. Govt. Print. Off., Washington, 1966.

\bibitem{Combis2012}
P.~Combis, P.~Cormont, L.~Gallais, D.~Hebert, L.~Robin, and J.~L. Rullier.
\newblock {Evaluation of the fused silica thermal conductivity by comparing
  infrared thermometry measurements with two-dimensional simulations}.
\newblock {\em {Appl. Phys. Lett.}}, 101:2--6, 2012.

\bibitem{Callister_2007}
W.~D. Callister~Jr.
\newblock {\em {Materials Science and Engineering: An Introduction}}.
\newblock John Wiley $\&$ Sons, Inc., New York, 7th edition, 2007.

\bibitem{Petrie2007}
E.~M. Petrie.
\newblock {\em {Handbook of Adhesives and Sealants, Second Edition}}.
\newblock McGraw-Hill Education, New York, 2nd edition, 2007.

\bibitem{Heraeus_2019}
Heraeus.
\newblock {\em Quartz Glass for Optics Data and Properties}, 2019.

\bibitem{siam}
L.~Kondic.
\newblock {Instability in the gravity driven flow of thin liquid films}.
\newblock {\em SIAM Review}, 45:95, 2003.

\bibitem{Zhang2010_GPU}
Y.~Zhang, J.~Cohen, and J.~D. Owens.
\newblock {Fast tridiagonal solvers on the GPU}.
\newblock {\em ACM SIGPLAN Notices}, 45:127--136, 2010.

\bibitem{Sanders}
J.~Sanders and E.~Kandrot.
\newblock {\em {CUDA by example}}.
\newblock Addison-Wesley, Upper Saddle River, NJ, 2011.

\bibitem{Gloster2019}
A.~Gloster, L.~Ó. Náraigh, and K.~E. Pang.
\newblock {cuPentBatch—A batched pentadiagonal solver for NVIDIA GPUs}.
\newblock {\em Comput. Phys. Commun}, 241:113--121, 2019.

\bibitem{GADIT_Thermal}
R.~H. Allaire.
\newblock Gadit thermal.
\newblock {https://github.com/Ryallaire/GADIT\_THERMAL}, 2021.

\end{thebibliography}
\bibliographystyle{unsrt}

\end{document}